\documentclass[preprint,showpacs,preprintnumbers,amsmath,amssymb]{revtex4}

\usepackage{graphicx}
\usepackage{dcolumn}
\usepackage{bm}

\begin{document}


\title{\boldmath
Two-parton twist-3 factorization in perturbative QCD
}
\vfill
\author{Makiko Nagashima$^1$}
\author{Hsiang-nan Li$^2$}%
%
\affiliation{$^1$Department of Physics, Ochanomizu University,\\
Bunkyo-ku, Tokyo 112-8610, Japan}
\affiliation{$^2$Institute of Physics, Academia Sinica,
Taipei, Taiwan 115, Republic of China}
\affiliation{$^2$Department of Physics, National Cheng-Kung
University,\\
Tainan, Taiwan 701, Republic of China
}%
%

%
%
\vfill
\begin{abstract}
We prove collinear factorization theorem for the process
$\pi\gamma^*\to\pi$ at the twist-3 level in the covariant gauge by
means of the Ward identity, concentrating on the two-parton case.
It is shown that soft divergences cancel and collinear divergences
are grouped into the pseudo-scalar and pseudo-tensor two-parton
twist-3 pion distribution amplitudes. The delicate summation of a
complete set of diagrams for achieving factorization in momentum,
spin, and color spaces is emphasized. The proof is then extended
to the exclusive semileptonic decay $B\to\pi l\bar\nu$, assuming
the hard scale to be of $O(\sqrt{\bar\Lambda M_B})$, where
$\bar\Lambda$ is a hadronic scale and $M_B$ the $B$ meson mass. We
explain the distinction between the factorization of collinear
divergences for a pion distribution amplitude and of soft
divergences for a $B$ meson distribution amplitude. The gauge
invariance and universality of the two-parton twist-3 pion
distribution amplitudes are confirmed. The proof presented here
can accommodate the leading twist-2 case. We then compare our
proof with that performed in the framework of soft-collinear
effective theory.

\end{abstract}
\maketitle 

\section{INTRODUCTION}

Recently, we have proposed a simple proof of collinear
factorization theorem in perturbative QCD (PQCD) for the exclusive
processes $\pi\gamma^*\to \gamma(\pi)$ and $B\to \gamma(\pi)
l\bar\nu$ based on the Ward identity \cite{L1}. According to this
theorem \cite{BL,BFL,MR,DM,CZS}, hadronic form factors are
factorized into the convolution of hard amplitudes with hadron
distribution amplitudes in momentum, spin, and color spaces. The
former, being infrared finite, are calculable in perturbation
theory. The latter, absorbing the infrared divergences involved in
the processes, are defined as matrix elements of nonlocal
operators. The universality of the distribution amplitudes and the
gauge invariance of the factorization have been explicitly
demonstrated. Our proof can be compared to that performed in the
axial gauge \cite{BL}, in which the factorization of infrared
divergences is trivial, but the gauge invariance is not obvious.
The formalism in \cite{L1} is restricted to the leading-twist,
{\it i.e.}, twist-2 level. As emphasized in \cite{L0,CKL},
contributions from the two-parton twist-3 pion distribution
amplitudes are not only chirally enhanced, but of the same power
as the leading-twist one in the semileptonic decay $B\to \pi
l\bar\nu$. Hence, it is necessary to derive the corresponding
factorization theorem. This proof can be regarded as an essential
step toward a rigorous construction of factorization theorem for
two-body nonleptonic $B$ meson decays.

The general decompositions of the matrix elements relevant to the
two-parton pion distribution amplitudes are, quoted from \cite{PB1},
\begin{eqnarray}
\langle 0|{\bar d}(y)\gamma_\mu \gamma_5u(0)|\pi^+(P)\rangle&=&
if_\pi P_{\mu}\int_0^1 dx e^{-ix P\cdot y}\phi_V(x)
+\frac{i}{2}f_\pi M_\pi^2\frac{y_\mu}{P\cdot y} \int_0^1 dx e^{-ix
P\cdot y}g_\pi(x)\;,
\label{pv}\\
\langle 0|{\bar d}(y)\gamma_5 u(0)|\pi^+(P)\rangle&=& -if_\pi
m_{0}\int_0^1 dx e^{-ix P\cdot y}\phi_S(x)\;,
\label{ps}\\
\langle 0|{\bar
d}(y)\gamma_5\sigma_{\mu\nu}u(0)|\pi^+(P)\rangle&=&
-\frac{i}{6}f_\pi m_{0}\left(1-\frac{M_\pi^2}{m_{0}^2}\right)
(P_{\mu}y_\nu-P_{\nu}y_\mu) \int_0^1 dx e^{-ix P\cdot
y}\phi_\sigma(x)\;, \label{pt}
\end{eqnarray}
where $\phi_{V,S,\sigma}$ and $g_\pi$ are the distribution
amplitudes of unit normalization, $f_\pi$ the pion decay constant,
$M_\pi$ the pion mass, $x$ the momentum fraction associated with
the $\bar d$ quark evaluated at the coordinate $y$. The Wilson
links that render the above nonlocal matrix elements gauge
invariant are not shown explicitly. It is easy to observe that the
contribution from $\phi_V$, independent of the pion mass, is
twist-2, and the contribution from $g_\pi$ is twist-4 because of
the factor $M_\pi^2$. The contributions from the pseudo-scalar
(PS) distribution amplitude $\phi_S$ and from the pseudo-tensor
(PT) distribution amplitude $\phi_\sigma$, proportional to the
chiral enhancing scale $m_{0}$, are twist-3.

We concentrate on the factorization of the two-parton twist-3
distribution amplitudes $\phi_S$ and $\phi_\sigma$ from the
processes $\pi\gamma^*\to\pi$ and $B\to\pi l\bar\nu$. We shall not
consider the three-parton twist-3 distribution amplitudes here,
since their contributions to exclusive processes are suppressed by
the strong coupling constant and of higher power. The reason is as
follows: after factorizing the corresponding infrared divergences
to all orders, the contribution is written as a convolution of a
hard amplitude with the three-parton twist-3 distribution
amplitudes. The hard amplitude contains one more attachment from
the extra parton (gluon) compared to that in the two-parton case.
The attachment introduces one more power of the coupling constant,
and one more hard propagator proportional to $1/Q$, where $Q$ is a
large scale characterizing the hard amplitude. Hence, a
three-parton hard amplitude is at least down by a power of the
coupling constant and a power of $1/Q$ compared to the
leading-order leading-twist hard amplitude. Moreover, the
three-parton twist-3 distribution amplitudes should be considered
along with the two-parton $k_T$ distribution amplitudes, which
form a complete gauge-invariant set.

Nonperturbative dynamics is reflected by infrared divergences of
radiative corrections in perturbation theory. There are two types
of infrared divergences, soft and collinear. Soft divergences come
from the region of a loop momentum $l$, where all its components
diminish. Collinear divergences are associated with a massless
quark of momentum $P\sim (Q,0,0_T)$. In the soft region and in the
collinear region with $l$ parallel to $P$, the components of $l$
behave like
\begin{eqnarray}
l^\mu=(l^+,l^-,l_T)\sim (\lambda,\lambda,\lambda)\;,\;\;\;\;
l^\mu\sim (Q,\lambda^2/Q,\lambda)\;,
\label{sog2}
\end{eqnarray}
respectively, where the light-cone coordinates have been adopted,
and $\lambda$ is a small hadronic scale. In both regions the
invariant mass of the radiated gluon diminishes as $\lambda^2$,
and the corresponding loop integrand may diverge as $1/\lambda^4$.
As the phase space for loop integration vanishes like $d^4 l\sim
\lambda^4$, logarithmic divergences are generated.

In this paper we shall derive the collinear factorization formula
for the scattering process $\pi\gamma^*\to\pi$, which involves the
pion form factor, at twist-3 by means of the Ward identity. The
chirally enhanced contributions to the pion form factor have been
calculated in \cite{GT} without proving their factorization
theorem. It will be shown that soft divergences cancel and
collinear divergences, factored out of the processes order by
order, are absorbed into the two-parton twist-3 pion distribution
amplitudes defined by the nonlocal matrix elements,
\begin{eqnarray}
\phi_S(x)&=&i\frac{P^+}{m_0}\int\frac{dy^-}{2\pi}e^{ixP^+y^-}
\langle 0|{\bar d}(y^-)\gamma_5 {\cal
P}\exp\left[-ig\int_0^{y^-}dzn_-\cdot A(zn_-)\right]u(0)|\pi^+(P)
\rangle\;,
\nonumber\\
\phi_T(x)&\equiv&
\frac{1}{6}\frac{d}{dx}\phi_\sigma(x)\;,\nonumber\\
&=&i\frac{P^+}{m_0}\int\frac{dy^-}{2\pi}e^{ixP^+y^-} \langle
0|{\bar d}(y^-)\gamma_5(\not\! n_+\not\! n_--1) {\cal
P}\exp\left[-ig\int_0^{y^-}dzn_-\cdot A(zn_-)\right]u(0)|\pi^+(P)
\rangle, \label{pwt}
\end{eqnarray}
where $n_+=(1,0,0_T)$ and $n_-=(0,1,0_T)$ are dimensionless
vectors on the light cone, the symbol $\cal P$ stands for
path-ordering of the Wilson line, and the pion decay constant
$f_\pi$ has been omitted. The definition of the hard amplitudes at
each order will be given as a result of the proof.

We then prove the collinear factorization theorem for the
semileptonic decay $B\to\pi l\bar\nu$, whose topology is similar
to the scattering process $\pi\gamma^*\to\pi$. In the heavy quark
limit the mass difference between the $B$ meson and the $b$ quark,
$\bar\Lambda=M_B-m_b$, represents a small scale. Assuming the hard
scale to be of $O(\sqrt{\bar\Lambda M_B})$, the soft divergences
do not cancel on the $B$ meson side, and the $B$ meson
distribution amplitudes are introduced to absorb the soft
divergences. The distinction between the factorization of soft
divergences for the $B$ meson distribution amplitudes and the
factorization of collinear divergences for the pion distribution
amplitudes will be explained. It will be shown that the two-parton
twist-3 pion distribution amplitudes derived from the scattering
$\pi\gamma^*\to \pi$ and from the decay $B\to\pi l\bar\nu$ are
identical as defined by Eq.~(\ref{pwt}). That is, the universality
of hadron distribution amplitudes is confirmed.

There are different opinions on whether the transverse degrees of
freedom of partons should be involved in exclusive $B$ meson
decays \cite{KPY,LPW,DC}. The conclusion drawn in \cite{LPW,DC}
that the parton transverse momenta $k_T$ are not necessary is
based on the analysis of the $B\to\gamma l\bar\nu$ decay, for
which the collinear factorization formula does not develop an
end-point singularity. When end-point singularities appear
\cite{SHB,BD,BF}, for example, in the collinear factorization
formulas of semileptonic and nonleptonic decays, the region with a
small momentum fraction $x$ becomes important. In this end-point
region the parton $k_T$ should be taken into account, and $k_T$
factorization theorem \cite{BS,LS} is more appropriate. Here we
shall derive the collinear factorization formalism for exclusive
$B$ meson decays. The $k_T$ dependence can be introduced
straightforwardly following the procedure in \cite{NL}.

We emphasize that the proof of factorization theorem is not the
whole story of PQCD. The double logarithms $\alpha_s\ln^2 x$
appearing in higher-order corrections to exclusive $B$ meson
decays have been observed \cite{KPY,LPW,ASY,L3,DS}. When the
end-point region is important, $\alpha_s\ln^2 x$ can not be
treated as a small expansion parameter, and should be summed to
all orders. A systematic treatment of these logarithms has been
proposed by grouping them into a quark jet function \cite{L2},
whose dependence on $x$ is governed by an evolution equation
\cite{L3}. A Sudakov factor, obtained by solving the evolution
equation, decreases fast at the end point. Moreover, if $k_T$
factorization theorem is adopted, $k_T$ resummation is also
required, which leads to another Sudakov factor describing the
parton distribution in $k_T$. Therefore, in a self-consistent
analysis the original factorization formulas should be convoluted
with the above two Sudakov factors. It turns out that the
end-point singularities do not exist \cite{TLS}, and an arbitrary
infrared cutoff for the momentum fraction $x$ \cite{SHB,BBNS} is
not required.

In the framework of soft-collinear effective theory (SCET)
\cite{bfl,bfps}, an effective Lagrangian with high-energy modes
integrated out has been constructed. This SCET Lagrangian provides
a simple guideline for deriving a factorization formula by
counting the powers of effective operators: start with an
effective operator relevant for a high-energy QCD process, and
draw the diagrams based on the SCET Lagrangians. Those effective
diagrams, whose contributions scale like the power the same as of
the operator, contribute to the matrix element formed by
the operator. This matrix element is identified as the
nonperturbative distribution amplitude, which collects the
infrared divergences in the process. The Wilson coefficient (the
hard amplitude) associated with the considered operator is then
obtained by subtracting the effective diagrams from the full
diagrams (the matching between the full theory and the effective
theory). An example for the application of SCET, the collinear
factorization of the $B\to D\pi$ decays, can be found in
\cite{CBP}. We shall compare the above formalism with ours
at the end of this paper. For a detailed comparison,
refer to \cite{Li03}.

We mention the opinion from the QCD-improved factorization (QCDF)
\cite{BBNS}, which claims that the $B\to\pi$ form factor,
suffering the end-point singularity in collinear factorization, is
dominated by soft dynamics. For a debate on this issue, refer to
\cite{DC,WY}. We have explained that the opposite conclusions on
the dominant dynamics in exclusive $B$ meson decays are attributed
to the different theoretical frameworks \cite{LL04}: a transition
form factor is factorizable in PQCD, i.e., in $k_T$ factorization as
explained above, not factorizable in QCDF, i.e., in collinear 
factorization (speaking of only the leading contribution) \cite{BBNS}, 
and partially factorizable in SCET \cite{BPS} (non-singular and 
singular pieces in collinear factorization are written into 
factorizable and nonfactorizable forms, respectively). Hence, there 
is no conflict at all among the above observations. It has been pointed
out that the collinear factorization and the $k_T$ factorization
lead to different phenomenological predictions for nonleptonic $B$
meson decays, such as the CP asymmetries in the $B\to\pi^+\pi^-$
modes \cite{NL,L03,KS02}. Therefore, it is expected that the
comparison with experimental data can discriminate the above
approaches.

In Sec.~II we derive the $O(\alpha_s)$ factorization of the
collinear divergences in the process $\pi\gamma^*\to\pi$. The
delicate summation of a complete set of diagrams for achieving the
factorizations in momentum, spin, and color spaces is emphasized.
The all-order proof based on the Ward identity is presented in
Sec.~III. The absence of the soft divergences is also shown. The
technique is then generalized to the decay $B\to\pi l\bar\nu$ in
Sec.~IV. The factorizations of soft divergences for the $B$ meson
distribution amplitudes and of collinear divergences for the pion
distribution amplitudes are compared. Section V is the conclusion. 
We refer the detailed calculations of the $O(\alpha_s)$ infrared
divergences in the above two processes to Appendices A, B, and C.

\section{$O(\alpha_s)$ FACTORIZATION of $\pi\gamma^*\to\pi$}

We start with the two-parton twist-3 factorization of the process
$\pi\gamma^*\to\pi$ at the one-loop level, which will serve as a
starting point of the all-order proof. The momentum $P_1$ ($P_2$)
of the initial-state (final-state) pion is parameterized as
\begin{eqnarray}
P_1=(P_1^+,0,{\bf 0}_T)=\frac{Q}{\sqrt{2}}(1,0,{\bf 0}_T)\;,
\;\;\;\; P_2=(0,P_2^-,{\bf 0}_T)=\frac{Q}{\sqrt{2}}(0,1,{\bf
0}_T)\;. \label{mpp}
\end{eqnarray}
Consider the kinematic region with large $Q^2=-q^2$, $q=P_2-P_1$
being the momentum transfer from the virtual photon, where PQCD is
applicable. The lowest-order diagrams with the valence quarks
being the external particles are displayed in Fig.~1. The lower
valence quark (an anti-quark ${\bar d}$) in the initial state pion
carries the fractional momentum $x_1P_1$. The lower valence quark
in the final state carries the fractional momentum $x_2P_2$.
Figure 1(a) gives the amplitude,
\begin{eqnarray}
G^{(0)}(x_1,x_2)=\frac{i}{2}eg^2 C_F
\frac{{\bar d}_i(x_1 P_1)I_{ij}
[\gamma^\nu d(x_2 P_2){\bar u}({\bar x}_2 P_2)
\gamma_\nu(\not\! P_2-x_1\not\! P_1)\gamma_\mu]_{jl}I_{lk}u_k({\bar x}_1P_1)}
{(P_2-x_1P_1)^2(x_1P_1-x_2P_2)^2}\;,
\label{h0}
\end{eqnarray}
with ${\bar x}_{1(2)}\equiv 1-x_{1(2)}$, the identity matrix $I$,
and the color factor $C_F$, where the averages over spins and
colors of the $u$ and ${\bar d}$ quarks have been done. The $u$
and $\bar d$ quark fields obey the equations of motion,
\begin{eqnarray}
{\bar x}_1\not\! P_1 u({\bar x}_1P_1)=0\;,\;\;\;\; {\bar
d}(x_1P_1)x_1\not\! P_1=0\;, \label{eqm}
\end{eqnarray}
where the quark masses $m_u$ and $m_d$ have been neglected in
high-energy scattering.

Insert the Fierz identity,
\begin{eqnarray}
I_{ij}I_{lk}&=&\frac{1}{4}I_{ik}I_{lj}
+\frac{1}{4}(\gamma_\alpha)_{ik}(\gamma^\alpha)_{lj}
+\frac{1}{4}(\gamma_5\not\! n_-)_{ik}(\not\! n_+\gamma_5)_{lj}
\nonumber\\
& &+\frac{1}{4}(\gamma_5)_{ik}(\gamma_5)_{lj}
+\frac{1}{4}[\gamma_5(\not\! n_+\not\! n_--1)]_{ik}
[(\not\! n_+\not\! n_--1)\gamma_5]_{lj}\;,
\label{1f}
\end{eqnarray}
into Eq.~(\ref{h0}) to separate the fermion flow. Different terms
in the above identity correspond to contributions of different
twists. The PS structure proportional to $\gamma_5$ and the PT
structure proportional to $\gamma_5(\not\! n_+\not\! n_--1)$ lead to
the twist-3 contributions. Equation (\ref{1f}) is a modified
version appropriate for extracting collinear divergences
\cite{TLS}: the choice of the PT structure in Eq.~(\ref{1f}) and
the ordinary one $(\gamma_5\sigma^{\alpha\beta})_{ik}
(\sigma_{\alpha\beta}\gamma_5)_{lj}$ are equivalent. The PS and PT
contributions to the process $\pi\gamma^*\to\pi$ must be included
simultaneously in order to form the gauge interaction vertex of a
pseudo-scalar particle, which is proportional to $(P_1+P_2)_\mu$.

The insertion on the initial-state side gives
\begin{eqnarray}
G^{(0)}(x_1,x_2)= \int d\xi_1\left[\phi_S^{(0)}(x_1,\xi_1)
H^{(0)}_{S}(\xi_1,x_2)+
\phi_T^{(0)}(x_1,\xi_1)H^{(0)}_{T}(\xi_1,x_2)\right]\;.
\end{eqnarray}
The functions $\phi_{S(T)}^{(0)}$ and $H_{S(T)}^{(0)}$,
\begin{eqnarray}
& &\phi_S^{(0)}(x_1,\xi_1)=\frac{1}{4m_0}{\bar d}(x_1P_1)\gamma_5
u({\bar x}_1P_1)\delta(\xi_1-x_1)\;,
\nonumber\\
& &\phi_T^{(0)}(x_1,\xi_1)=\frac{1}{4m_0}{\bar d}(x_1P_1)
\gamma_5(\not\! n_+\not\! n_--1)u({\bar x}_1P_1)\delta(\xi_1-x_1)\;,
\nonumber\\
& &H_{S}^{(0)}(\xi_1,x_2)=\frac{i}{2}eg^2 C_Fm_0\frac{tr[\gamma^\nu
d(x_2 P_2){\bar u}({\bar x}_2 P_2)
\gamma_\nu(\not\! P_2-\xi_1\not\! P_1)
\gamma_\mu\gamma_5]}{(P_2-\xi_1P_1)^2(\xi_1P_1-x_2P_2)^2}\;,
\nonumber\\
& &H_{T}^{(0)}(\xi_1,x_2)=\frac{i}{2}eg^2 C_Fm_0\frac{tr[\gamma^\nu
d(x_2 P_2){\bar u}({\bar x}_2 P_2)
\gamma_\nu(\not\! P_2-\xi_1\not\! P_1)
\gamma_\mu(\not\! n_+\not\! n_--1)\gamma_5]}{(P_2-\xi_1P_1)^2
(\xi_1P_1-x_2P_2)^2}\;,
\label{low}
\end{eqnarray}
define the lowest-order perturbative PS (PT) distribution
amplitude and the corresponding hard amplitude, respectively. The
meaning of the variable $\xi_1$, regarded as a momentum fraction
modified by collinear gluon exchanges, will become clear below.

Consider the $O(\alpha_s)$ radiative corrections to Fig.~1(a) in
the covariant (Feynman) gauge, which are shown in Fig.~2, and
identify their infrared divergences. Self-energy corrections to
the internal lines, leading to a next-to-leading-order hard
amplitude, are not included. Here we summarize only the results of
the $O(\alpha_s)$ factorization, and leave the details to Appendix
A. It will be shown that all the diagrams in Fig.~2 can be written
as the convolution of the lowest-order hard amplitudes
$H_{S,T}^{(0)}$ in Eq.~(\ref{low}) with the $O(\alpha_s)$
divergent distribution amplitudes $\phi_{S,T}^{(1)}$ in the
collinear region with the loop momentum $l$ parallel to $P_1$. The
expressions of $\phi_{S,T}^{(1)}$ will provide a basis of
constructing the all-order definitions of the two-parton twist-3
distribution amplitudes as nonlocal matrix elements.

Figures 2(a)-2(c) are the two-particle reducible diagrams with the
additional gluon attaching the two valence quarks of the initial
state. It has been known that soft divergences cancel among these
diagrams. The reason for this cancellation is that soft gluons,
being huge in space-time, do not resolve the color structure of
the pion. Collinear divergences in Figs.~2(a)-2(c) do not cancel,
since the loop momentum flows into the internal lines in
Fig.~2(b), but not in Figs.~2(a) and 2(c). Inserting the Fierz
identity into Figs.~2(a)-2(c), we obtain the approximate loop
integrals in the collinear region,
\begin{eqnarray}
I^{(a),(b),(c)}&\approx &\sum_{n=S,T} \int
d\xi_1\phi^{(1)}_{na,nb,nc}(x_1,\xi_1) H_n^{(0)}(\xi_1,x_2)\;,
\label{2b2}
\end{eqnarray}
respectively. The $O(\alpha_s)$ PS pieces,
\begin{eqnarray}
\phi^{(1)}_{Sa}(x_1,\xi_1)&=& \frac{-g^2 C_F}{8m_0}
\int\frac{d^4l}{(2\pi)^4}{\bar d}(x_1P_1)
\gamma_5\frac{i}{\bar{x}_1\not\! P_1}
\gamma_\beta\frac{{\bar x}_1\not\! P_1 +\not l}
{({\bar x}_1 P_1 +l)^2}\gamma^\beta u({\bar x}_1P_1)\frac{1}{l^2}
\delta(\xi_1-x_1)\;,
\label{p2a}\\
\phi^{(1)}_{Sb}(x_1,\xi_1)&=& \frac{ig^2 C_F}{4m_0}
\int\frac{d^4l}{(2\pi)^4}{\bar d}(x_1P_1)\gamma_\beta
\frac{x_1\not\! P_1-\not l}{(x_1P_1-l)^2}\gamma_5 \frac{{\bar
x}_1\not\! P_1 +\not l}{({\bar x}_1 P_1 +l)^2}\gamma^\beta u({\bar
x}_1P_1)\frac{1}{l^2}
\delta\left(\xi_1-x_1+\frac{l^+}{P_1^+}\right),
\label{p2b}\\
\phi^{(1)}_{Sc}(x_1,\xi_1)&=& \frac{g^2 C_F}{8m_0}
\int\frac{d^4l}{(2\pi)^4}{\bar d}(x_1P_1)\gamma_\beta
\frac{x_1\not\! P_1-\not l}{(x_1P_1-l)^2}\gamma^\beta
\frac{-i}{x_1\not\! P_1}
\gamma_5u({\bar x}_1P_1)\frac{1}{l^2}\delta(\xi_1-x_1)\;,
\label{p2c}
\end{eqnarray}
and the PT pieces with $\gamma_5$ in the above expressions being
replaced by $\gamma_5(\not\! n_+\not\! n_--1)$, contain the collinear
(logarithmic) divergences in Figs.~2(a), 2(b), and 2(c),
respectively. Note that the momentum fraction $x_1$ in Fig.~2(b)
has been modified into $\xi_1=x_1-l^+/P_1^+$, because the loop
momentum $l$ flows through the hard gluon.

The factorization of the loop integrals associated with the two-particle
irreducible diagrams in Fig.~2(d)-2(f) is written as
\begin{eqnarray}
I^{(d),(e),(f)}&\approx& \sum_{n=S,T}\int d\xi_1
\phi_{nd,ne,nf}^{(1)}(x_1,\xi_1)H_n^{(0)}(\xi_1,x_2)\;,
\label{2f2}
\end{eqnarray}
with the $O(\alpha_s)$ PS pieces,
\begin{eqnarray}
\phi_{Sd}^{(1)}(x_1,\xi_1)&=&\frac{-ig^2}{2m_0C_F}
\int\frac{d^4l}{(2\pi)^4}{\bar d}(x_1P_1)\gamma_5
\frac{{\bar x}_1\not\! P_1 +\not l}{({\bar x}_1 P_1 +l)^2}
\gamma^\beta u({\bar x}_1P_1)\frac{1}{l^2}
\frac{n_{-\beta}}{n_-\cdot l}
\nonumber\\
& &\times\left[\delta(\xi_1-x_1)-
\delta\left(\xi_1-x_1+\frac{l^+}{P_1^+}\right)\right]\;,
\label{p2d}\\
\phi_{Se}^{(1)}(x_1,\xi_1)
&=&\frac{ig^2}{8m_0N_c}
\int\frac{d^4l}{(2\pi)^4}{\bar d}(x_1P_1)\gamma_5
\frac{{\bar x}_1\not\! P_1 +\not l}{({\bar x}_1 P_1 +l)^2}\gamma^\beta
u({\bar x}_1P_1)\frac{1}{l^2}
\frac{n_{-\beta}}{n_-\cdot l}\delta(\xi_1-x_1)\;,
\label{p2e}\\
\phi_{Sf}^{(1)}(x_1,\xi_1)&=&\frac{-ig^2}{8m_0N_c}
\int\frac{d^4l}{(2\pi)^4}{\bar d}(x_1P_1)\gamma_5
\frac{{\bar x}_1\not\! P_1 +\not l}{({\bar x}_1 P_1 +l)^2}\gamma^\beta
u({\bar x}_1P_1)\frac{1}{l^2}
\frac{n_{-\beta}}{n_-\cdot l}
\delta\left(\xi_1-x_1+\frac{l^+}{P_1^+}\right)\;,
\label{p2f}
\end{eqnarray}
$N_c=3$ being the number of colors. The corresponding PT pieces
are defined similarly with $\gamma_5$ in the above expressions
being replaced by $\gamma_5(\not\! n_+\not\! n_--1)$. Figure 2(g) does
not contain a collinear divergence. Note that Fig.~2(d) is free of
a soft divergence, because the additional gluon attaches the hard
gluon. The soft divergences cancel between Figs.~2(e) and 2(f).
The above absence of the soft divergences is obvious from the
cancellation in Eq.~(\ref{p2d}) and between Eqs.~(\ref{p2e}) and
(\ref{p2f}) as $l\to 0$. The contribution from Fig.~2(d) has been
split into two terms as a consequence of the Ward identity
\cite{L1}. The first and second $\delta$-functions in
Eq.~(\ref{p2d}) correspond to the cases without and with the loop
momentum $l$ flowing through the internal lines, respectively. The
Feynman rule $n_{-\beta}/n_-\cdot l$, coming from the eikonal
approximation (see Appendix A), could be generated by a Wilson
line in the direction of $n_-$. This factor will not appear, if
the proof is performed in the axial gauge $n_-\cdot A=0$.

Note that Eqs.~(\ref{p2d})-(\ref{p2f}) possess
different color factors due to different color flows in Figs.~2(d)-2(f).
Combining the contributions from Figs.~2(d)-2(f), we arrive at
\begin{eqnarray}
\sum_{i=(d)}^{(f)}I^{i}&\approx&\sum_{n=S,T}\int d\xi_1
\phi_{nu}^{(1)}(x_1,\xi_1)H_n^{(0)}(\xi_1,x_2)\;,
\label{2dg}
\end{eqnarray}
where the PS piece,
\begin{eqnarray}
\phi_{Su}^{(1)}(x_1,\xi_1)&=&\frac{-ig^2C_F}{4m_0}
\int\frac{d^4l}{(2\pi)^4}{\bar d}(x_1P_1)\gamma_5
\frac{{\bar x}_1\not\! P_1 +\not l}{({\bar x}_1 P_1 +l)^2}\gamma^\beta
u({\bar x}_1P_1)\frac{1}{l^2}
\frac{n_{-\beta}}{n_-\cdot l}
\nonumber\\
& &\times\left[\delta(\xi_1-x_1)-
\delta\left(\xi_1-x_1+\frac{l^+}{P_1^+}\right)\right]\;,
\label{2nu}
\end{eqnarray}
is associated with the collinear gluon emitted from the $u$ quark.
The color factor $C_F$ implies the factorization of the
distribution amplitude from other parts of the process in color
space, which can be achieved only by summing a set of diagrams.
The first and second terms in Eq.~(\ref{2nu}) correspond to
Figs.~3(a) and 3(b), respectively, where the double lines
represent the Wilson lines mentioned above.

The analysis of Figs.~2(h)-2(k) is similar, and the conclusion is that
Fig.~2(h) is split into two terms as in Eq.~(\ref{p2d}), the collinear
gluons in Figs.~2(i) and 2(j) are eikonalized as in Eqs.~(\ref{p2e}) and
(\ref{p2f}), respectively, and Fig.~2(k), like Fig.~2(g), does not
contribute a collinear divergence. The soft divergences cancel among the
above diagrams. Summing over Figs.~2(h)-2(j), we derive the correct color
factor:
\begin{eqnarray}
\sum_{i=(h)}^{(j)}I^{i}&\approx&\sum_{n=S,T}\int d\xi_1
\phi_{n\bar d}^{(1)}(x_1,\xi_1)H_n^{(0)}(\xi_1,x_2)\;,
\label{2hk}
\end{eqnarray}
where the PS piece,
\begin{eqnarray}
\phi_{S\bar d}^{(1)}(x_1,\xi_1)&=&\frac{ig^2C_F}{4m_0}
\int\frac{d^4l}{(2\pi)^4}{\bar d}(x_1P_1)
\gamma^\beta\frac{x_1\not\! P_1 -\not l}{(x_1 P_1 -l)^2}
\gamma_5 u({\bar x}_1P_1)\frac{1}{l^2}
\frac{n_{-\beta}}{n_-\cdot l}
\nonumber\\
& &\times \left[\delta(\xi_1-x_1)-
\delta\left(\xi_1-x_1+\frac{l^+}{P_1^+}\right)\right]\;,
\label{2nd}
\end{eqnarray}
is associated with the collinear gluon emitted from the $\bar d$
quark. The first and second terms in Eq.~(\ref{2nd}) correspond to
Figs.~3(c) and 3(d), respectively.

The sum of Eqs.~(\ref{2b2}), (\ref{2dg}) and (\ref{2hk})
leads to
\begin{eqnarray}
\sum_{i=(a)}^{(k)}I^{i}\approx\sum_{n=S,T}\int d\xi_1
\phi_n^{(1)}(x_1,\xi_1)H_n^{(0)}(\xi_1,x_2)\;,
\label{phi1}
\end{eqnarray}
where $\phi_S^{(1)}$ and $\phi_T^{(1)}$ are represented by the
$O(\alpha_s)$ terms of the nonlocal matrix elements with
the structures $\gamma_5$ and $\gamma_5(\not\! n_+\not\! n_--1)$ sandwiched,
\begin{eqnarray}
\phi_{S}(x,\xi)&=&i\frac{P^+}{m_0}\int \frac{dy^-}{2\pi}e^{i\xi
P^+y^-} \langle 0|{\bar d}(y^-)\gamma_5{\cal P}\exp\left[-ig
\int_0^{y^-}dzn_-\cdot A(zn_-)\right]u(0)|u(\bar xP)\bar
d(xP)\rangle\;,
\nonumber\\
\phi_{T}(x,\xi)&=&i\frac{P^+}{m_0}\int \frac{dy^-}{2\pi}e^{i\xi P^+y^-}
\langle 0|{\bar d}(y^-)\gamma_5(\not\! n_+\not\! n_--1)
\nonumber\\
& &\times {\cal P}\exp\left[-ig \int_0^{y^-}dzn_-\cdot
A(zn_-)\right]u(0)|u(\bar xP)\bar d(xP)\rangle\;, \label{ld}
\end{eqnarray}
respectively. By expanding the quark field ${\bar d}(y^-)$ and the
path-ordered exponential (Wilson line) into powers of $y^-$, the
above matrix elements can be expressed as a series of the
covariant derivatives $(D^+)^n{\bar d}(0)$, and are gauge
invariant. The integral over $z$ in fact contains two pieces: for
the upper Wilson line in Fig.~3(a), $z$ runs from 0 to $\infty$.
For the lower Wilson line in Fig.~3(b), $z$ runs from $\infty$
back to $y^-$. The light-cone coordinate $y^-\not =0$ corresponds
to the fact that the collinear divergences in Fig.~2 do not
cancel. Notice the different kets $|\pi^+(P) \rangle$ and $|u(\bar
xP)\bar d(xP)\rangle$ in Eqs.~(\ref{pwt}) and (\ref{ld}),
respectively. Equation~(\ref{ld}) plays the role of an infrared
regulator for parton-level diagrams. A hard amplitude, obtained by
subtracting Eq.~(\ref{ld}) from the parton-level diagrams, then
corresponds to the regularized parton-level diagrams. After
determining the gauge-invariant infrared-finite hard amplitude
$H(x)$, we convolute it with the physical two-parton twist-3 pion
distribution amplitudes defined in Eq.~(\ref{pwt}) and graphically
shown in Fig.~3(e). Models for the two-parton twist-3 pion
distribution amplitudes have been derived using QCD sum rules
\cite{PB1}.

It is easy to confirm that $\phi_{S,T}^{(1)}$ is reproduced by the
perturbative expansion of the matrix elements in Eq.~(\ref{ld}).
Take Eq.~(\ref{2nu}) as an example. Fourier transforming the gauge
field $A(zn_-)$ into ${\tilde A}(l)$, we have
\begin{eqnarray}
-ig\int_0^{\infty}dz \exp[iz(n_-\cdot l+i\epsilon)]n_-\cdot
{\tilde A}(l) =g\frac{n_-^\alpha}{n_-\cdot l} {\tilde
A}_\alpha(l)\;.
\end{eqnarray}
The field ${\tilde A}_\alpha(l)$, contracted with another gauge
field associated with the $u$ quark, gives the gluon propagator in
Fig.~3(a). It is then realized that the eikonal propagator is
generated by the path-ordered exponential. The second piece of the
Wilson line corresponds to the second term in Eq.~(\ref{2nu}):
\begin{eqnarray}
-ig\int_{\infty}^{y^-}dz \exp[iz(n_-\cdot l+i\epsilon)]n_-\cdot
{\tilde A}(l) =-g\frac{n_-^\alpha}{n_-\cdot l} \exp(i
l^+y^-){\tilde A}_\alpha(l)\;,
\end{eqnarray}
where the extra Fourier factor $\exp(i l^+y^-)$
leads to the function $\delta(\xi-x+l^+/P_1^+)$.
The field ${\tilde A}_\alpha(l)$, contracted with another gauge field
associated with the $u$ quark, gives the gluon propagator in Fig.~3(b).
The Feynman rules for Eq.~(\ref{2nd}) can be reproduced in a similar way,
where ${\tilde A}_\alpha(l)$ is contracted with another gauge field
associated with the $\bar d$ quark.
Equations (\ref{p2a})-(\ref{p2c}) are derived by contracting the
gluon fields associated with the $u$ and $\bar d$ quarks.

The above derivation also applies straightforwardly to the
factorization of the collinear divergences associated with the
final state, which arise from the region with the loop momentum
parallel to $P_2$. Hence, we have an expression similar to
Eq.~(\ref{phi1}), $\sum_{m=S,T}H^{(0)}_m\otimes\phi^{(1)}_m$,
where $\phi_m^{(1)}$ is the final-state distribution amplitude,
and $H_m^{(0)}$ is the hard amplitude with the Fierz identity
inserted into the final-state side of $H^{(0)}$. The symbol
$\otimes$ represents the convolution in the variable $\xi_2$. The
momentum fraction associated with the final state will be modified
into $\xi_2=x_2-l^-/P_2^-$ by the collinear gluons parallel to
$P_2$, if the loop momentum $l$ flows through the hard amplitude
$H_m^{(0)}$.

The $O(\alpha_s)$ radiative corrections to Fig.~1(b) are displayed
in Fig.~4. The factorization of the collinear divergences from
these diagrams is referred to Appendix B. The result is similar to
Eq.~(\ref{phi1}) but without the PT contributions, because of
$\gamma^\nu (\not\! n_+\not\! n_--1)\gamma_\nu=0$, where the gamma
matrices $\gamma^\nu$ and $\gamma_\nu$ come from the gluon
vertices in Fig.~1(b). Hence, we derive, from Fig.~4,
\begin{eqnarray}
\sum_{i=(a)}^{(k)}I^{i}\approx\int d\xi_1
\phi_S^{(1)}(x_1,\xi_1)H_S^{(0)}(\xi_1,x_2)\;, \label{phi3}
\end{eqnarray}
where the definition of the $O(\alpha_s)$ PS distribution
amplitude $\phi_S^{(1)}$ is the same as in Eq.~(\ref{ld}). This is
expected due to the universality.

In summary, the $O(\alpha_s)$
factorization of the process $\pi\gamma^*\to\pi$ is written as
\begin{eqnarray}
G^{(1)}=\sum_{n=S,T}\phi^{(1)}_n\otimes H^{(0)}_n
+\sum_{m=S,T}H^{(0)}_m\otimes\phi^{(1)}_m+H^{(1)}\;,
\label{fac1}
\end{eqnarray}
where $G^{(1)}$ denotes the complete set of the $O(\alpha_s)$
corrections, and $H_{n,m}^{(0)}$ receive the contributions from
both Figs.~1(a) and 1(b) now. The first term on the right-hand
side of the above expression does not contain the collinear
divergences from the loop momentum $l$ parallel to $P_2$. In this
region $l^+$ is negligible, $\xi_1$ approaches $x_1$, and the
corresponding collinear divergences cancel in $\phi_n^{(1)}$. For
the similar reason, the second term on the right-hand side of
Eq.~(\ref{fac1}) does not contain the collinear divergences from
$l$ parallel to $P_1$. That is, the initial-state and final-state
collinear divergences in $G^{(1)}$ have been completely factorized
into the first and second terms on the right-hand side of
Eq.~(\ref{fac1}), respectively. The $O(\alpha_s)$ hard amplitude
$H^{(1)}$, defined via Eq.~(\ref{fac1}), is infrared finite. Note
that $H^{(1)}$ contains the contributions from Figs.~2(g), 2(k)
and the self-energy corrections to the internal lines.

Summing Eq.~(\ref{fac1}) and the lowest-order diagrams
$G^{(0)}$, the factorization formula for the
two-parton twist-3 contributions to the process $\pi\gamma^*\to\pi$ is
given, up to $O(\alpha_s)$, by
\begin{eqnarray}
G^{(0)}+G^{(1)}=\sum_{n,m=S,T}(\phi^{(0)}_n+\phi^{(1)}_n)
\otimes(H_{nm}^{(0)}+H^{(1)}_{nm})\otimes
(\phi^{(0)}_m+\phi^{(1)}_m)\;,
\label{ff1}
\end{eqnarray}
where the trivial factorizations,
\begin{eqnarray}
H_{n}^{(0)}=\sum_{m=S,T}H_{nm}^{(0)}\otimes \phi_{m}^{(0)}\;,
\;\;\;\;
H_{m}^{(0)}=\sum_{n=S,T}\phi_{n}^{(0)}\otimes H_{nm}^{(0)}\;,
\;\;\;\;
H^{(1)}=\sum_{n,m=S,T}\phi_n^{(0)}\otimes H_{nm}^{(1)}
\otimes \phi^{(0)}_m\;,
\label{wi3}
\end{eqnarray}
have been adopted. The last formula in Eq.~(\ref{wi3}) defines the 
$O(\alpha_s)$ hard amplitude $H^{(1)}_{nm}$.The explicit expression 
of, for example, $H^{(0)}_{SS}$, is given by
\begin{eqnarray}
H_{SS}^{(0)}(\xi_1,\xi_2)=\frac{i}{2}eg^2
C_Fm_0^2\Bigg\{\frac{tr[\gamma^\nu \gamma_5 \gamma_\nu(\not\!
P_2-\xi_1\not\! P_1)\gamma_\mu\gamma_5]}{(P_2-\xi_1P_1)^2
(\xi_1P_1-\xi_2P_2)^2}+\frac{tr[\gamma^\nu \gamma_5
\gamma_\mu(\not\! P_1-\xi_2\not\!
P_2)\gamma_\nu\gamma_5]}{(P_1-\xi_2P_2)^2
(\xi_1P_1-\xi_2P_2)^2}\Bigg\}\;. \label{ss0}
\end{eqnarray}
It is obvious that the PS and PT
structures must be included simultaneously for a complete
two-parton twist-3 collinear factorization.

\section{ALL-ORDER FACTORIZATION of $\pi\gamma^*\to\pi$}

In this section we present the all-order proof of two-parton
twist-3 factorization theorem for the process $\pi\gamma^*\to
\pi$, and construct the gauge-invariant distribution amplitudes in
Eq.~(\ref{ld}). It has been mentioned in the Introduction that
factorizations of a QCD process in momentum, spin, and color
spaces require summation of many diagrams, especially at higher
orders. The diagram summation can be handled in an elegant way by
employing the Ward identity,
\begin{eqnarray}
l^\mu G_\mu(l,k_1,k_2,\cdots, k_n)=0\;, \label{war}
\end{eqnarray}
where $G_\mu$ represents a physical amplitude with an external
gluon carrying the momentum $l$ and with $n$ external quarks
carrying the momenta $k_1$, $k_2$, $\cdots$, $k_n$. All these
external particles are on mass shell. The Ward identity can be
easily derived by means of the Becchi-Rouet-Stora (BRS)
transformation \cite{BRS}. We shall employ the similar Ward
identity,
\begin{eqnarray}
l_{1}^{\mu} G_{\mu\nu}(l_1,l_2,k_1,k_2,\cdots,
k_n)l_{2}^{\nu}=0\;, \label{war2}
\end{eqnarray}
where $G_{\mu\nu}$ represents a physical amplitude with two
external gluons carrying the momenta $l_1$ and $l_2$, and with $n$
external quarks.

We shall prove two-parton twist-3 factorization theorem for the process
$\pi\gamma^*\to\pi$ to all orders by induction. The factorization of the
$O(\alpha_s)$ collinear divergences has been worked out in Sec.~II.
Assume that factorization theorem holds up to $O(\alpha_s^N)$:
\begin{eqnarray}
G^{(k)}=\sum_{n',m'=S,T}\sum_{i=0}^{k}\sum_{j=0}^{k-i}
\phi_{n'}^{(i)}\otimes
H_{n'm'}^{(k-i-j)}\otimes \phi_{m'}^{(j)}\;,\;\;\;k=0,1,\cdots N\;.
\label{gnf}
\end{eqnarray}
$G^{(k)}$ denotes the parton-level diagrams of $O(\alpha_s^k)$
with $G^{(0)}$ shown in Fig.~1. The initial-state distribution
amplitude $\phi_{n'}^{(i)}$ is defined by the $O(\alpha_s^i)$
terms in the perturbative expansion of Eq.~(\ref{ld}), and the
final-state distribution amplitude $\phi_{m'}^{(j)}$ defined
similarly by the complex conjugate of Eq.~(\ref{ld}).
$H_{n'm'}^{(k-i-j)}$ is the remaining $O(\alpha_s^{k-i-j})$ piece
of the process, which does not contain the infrared divergences.
Equation (\ref{gnf}) implies that all the initial-state and
final-state collinear divergences in $G^{(k)}$ have been collected
into $\phi_{n'}^{(i)}$ and $\phi_{m'}^{(j)}$ systematically.
Inserting the Fierz identity, we also obtain the trivial
factorizations of the distribution amplitudes $\phi$ and the
diagrams $G$ at arbitrary orders of $\alpha_s$, similar to
Eq.~(\ref{wi3}). We then have the factorization,
\begin{eqnarray}
G_{nm}^{(k)}=\sum_{n',m'=S,T}\sum_{i=0}^{k}\sum_{j=0}^{k-i}
\phi_{nn'}^{(i)}\otimes H_{n'm'}^{(k-i-j)}\otimes
\phi_{m'm}^{(j)}\;, \label{gnf1}
\end{eqnarray}
in which, for example, $\phi_{SS}^{(1)}$ contains the piece
$\phi_{SSu}^{(1)}$ extracted from Eq.~(\ref{2nu}),
\begin{eqnarray}
\phi_{SSu}^{(1)}(x_1,\xi_1)&=&\frac{-ig^2C_F}{4m_0}
\int\frac{d^4l}{(2\pi)^4}tr\left[\gamma_5 \frac{{\bar x}_1\not\! P_1
+\not l}{({\bar x}_1 P_1 +l)^2}\gamma^\beta
\gamma_5\right]\frac{1}{l^2} \frac{n_{-\beta}}{n_-\cdot l}
\nonumber\\
& &\times\left[\delta(\xi_1-x_1)-
\delta\left(\xi_1-x_1+\frac{l^+}{P_1^+}\right)\right]\;,
\label{3nu}
\end{eqnarray}

Below we prove the collinear factorization of the
$O(\alpha_s^{N+1})$ diagrams $G^{(N+1)}$, assuming Eq.~(\ref{gnf})
or (\ref{gnf1}). Look for the radiative gluon in a subset of
$O(\alpha_s^{N+1})$ diagrams $G^{(N+1)}$, one of whose ends
attaches the outer most vertex on the upper $u$ quark line. Let
$\alpha$ denote the outer most vertex, and $\beta$ denote the
attachments of the other end of the identified gluon inside the
rest of the diagrams. There are two types of collinear
configurations associated with this gluon, depending on whether
the vertex $\beta$ is located on an internal line with the
momentum along $P_1$. The fermion propagator adjacent to the
vertex $\alpha$ is proportional to $\not\! P_1$ in the collinear
region with the loop momentum $l$ parallel to $P_1$. If $\beta$ is
not located on a collinear line along $P_1$, the component
$\gamma^+$ in $\gamma^\alpha$ and the minus component of the
vertex $\beta$ give the leading contribution. If $\beta$ is
located on a collinear line along $P_1$, $\beta$ can not be minus,
and both $\alpha$ and $\beta$ represent the transverse components.
This configuration is the same as of the self-energy correction to
an on-shell particle. According to the above classification, we
decompose the tensor $g_{\alpha\beta}$ appearing in the propagator
of the identified gluon into \cite{L1},
\begin{eqnarray}
g_{\alpha\beta}=\delta_{\alpha +}\delta_{\beta -}
-\delta_{\alpha \perp}\delta_{\beta \perp}
+\delta_{\alpha -}\delta_{\beta +}\;.
\label{dec}
\end{eqnarray}
The first (second) term on the right-hand side of Eq.~(\ref{dec})
extracts the first (second) type of initial-state collinear divergences
mentioned above. The third term does not
contribute due to the equations of motion in Eq.~(\ref{eqm}).

We discuss the factorization of the first type of collinear
configurations denoted by $G_I^{(N+1)}$. As stated before, the
identified collinear gluon with $\alpha=+$ and $\beta=-$ does not
attach the upper or lower quark line directly, which carries the
momentum along $P_1$. That is, those diagrams with Figs.~2(a) and
2(b) as the $O(\alpha_s)$ subdiagrams are excluded from the set of
$G^{(N+1)}$ as discussing the first type of collinear
configurations. We employ the replacement,
\begin{eqnarray}
\delta_{\alpha +}\delta_{\beta -}\to
\frac{n_{-\alpha} l_\beta}{n_-\cdot l}\;,
\label{rep}
\end{eqnarray}
where the light-like vector $n_{-\alpha}$ also selects the plus
component of $\gamma^\alpha$, and $l_\beta$ selects the minus
component of the vertex $\beta$ in the collinear region. $l_\beta$
can attach all the internal lines, no matter they are or are not
parallel to $P_1$. When it attaches a line parallel to $P_1$, the
result diminishes. That is, Eq.~(\ref{rep}) indeed picks up the
first type of collinear configurations. We then consider
Fig.~5(a), which contains a complete set of contractions of
$l_\beta$, since the second and third diagrams have been included.
The contractions of $l_\beta$, represented by arrows, hint the
application of the Ward identity in Eq.~(\ref{war}) to the case,
in which the on-shell external $u$ quark, $\bar d$ quark and gluon
carry the momenta ${\bar \xi}_1 P_1$, $x_1P_1$ and $l$,
respectively. The Ward identity states that the expression in
Fig.~5(a) vanishes. The second and third diagrams in Fig.~5(a)
give
\begin{eqnarray}
& &l_\beta \frac{1}{{\bar \xi}_1\not\! P_1-\not l}\gamma^\beta
u({\bar \xi}_1 P_1) =\frac{1}{{\bar \xi}_1\not\! P_1-\not l}(\not
l-{\bar \xi}_1\not\! P_1 + {\bar \xi}_1\not\! P_1)u({\bar \xi}_1 P_1)
=-u({\bar \xi}_1 P_1)\;,
\nonumber\\
& &l_\beta{\bar d}(x_1P_1)\gamma^\beta\frac{1}{x_1\not\! P_1-\not l}
=-{\bar d}(x_1P_1)\;,
\label{ide2}
\end{eqnarray}
respectively, where the equations of motion in Eq.~(\ref{eqm})
have been employed. The terms $u({\bar \xi}_1 P_1)$ and ${\bar
d}(x_1P_1)$ at the ends of the above expressions are associated
with the $O(\alpha_s^N)$ diagrams.

We then insert the Fierz identity into Fig.~5(a), and factor the
lowest-order expressions ${\bar d}(x_1P_1)\Gamma u({\bar \xi}_1
P_1)$ with $\Gamma$ being the PS or PT structure considered in
this work. The result is a relation shown in Fig.~5(b), where the
cuts on the quark lines denote the insertion of the Fierz
identity, and the double (Wilson) lines represent
$n_{-\alpha}/n_-\cdot l$ in Eq.~(\ref{rep}). Figure 5(b) implies
that the diagrams $G_I^{(N+1)}$ corresponding to the first term in
Eq.~(\ref{dec}) are factorized into the convolution of the full
diagrams $G_n^{(N)}$ with $\phi_{nu}^{(1)}$, $n=S,T$. The same
discussion applies to the factorization of the diagrams
$G_I^{(N+1)}$ with the collinear gluon emitted from the outer most
vertex on the ${\bar d}$ quark line, leading to a convolution of
$G_n^{(N)}$ with $\phi_{n\bar d}^{(1)}$. Similarly, for the subset
of $G^{(N+1)}$, in which the identified radiative gluon emitted by
the outgoing quarks, the collinear divergences are also classified
into two types. In this case it is the third term in
Eq.~(\ref{dec}) that corresponds to the first type of collinear
divergences, and the replacement in Eq.~(\ref{rep}) is modified
into
\begin{eqnarray}
\delta_{\alpha -}\delta_{\beta +}\to
\frac{n_{+\alpha} l_\beta}{n_+\cdot l}\;.
\label{rep1}
\end{eqnarray}
The pieces $\phi_{mu}^{(1)}$ and $\phi_{m\bar d}^{(1)}$ containing
the identified gluon are factored out of $G_I^{(N+1)}$ in the
collinear region with the loop momentum parallel to $P_2$.

We conclude that $G^{(N+1)}_{\parallel\cdot}$, in which the
replacement in Eq.~(\ref{rep}) is applicable to the initial-state
side, and $G^{(N+1)}_{\cdot\parallel}$, in which the replacement
in Eq.~(\ref{rep1}) is applicable to the final-state side, are
written as,
\begin{eqnarray}
G^{(N+1)}_{\parallel\cdot}&\approx& \sum_{n=S,T}
\phi_{n\parallel}^{(1)}\otimes G_n^{(N)}\;,
\label{wic}\\
G^{(N+1)}_{\cdot\parallel}&\approx& \sum_{m=S,T} G_m^{(N)}\otimes
\phi_{m\parallel}^{(1)}\;,
\label{wicr}
\end{eqnarray}
respectively, where $\phi_{n(m)\parallel}^{(1)}$ represents
\begin{eqnarray}
\phi_{n(m)\parallel}^{(1)}=\phi_{n(m)u}^{(1)}+\phi_{n(m)\bar d}^{(1)}\;.
\end{eqnarray}
Equation (\ref{wic}) is displayed in Fig.~6. Other diagrams, which
do not contain the radiative gluons on either the initial-state
side or the final-state side, are self-energy corrections to
internal lines. They are infrared finite, and contribute to the
$O(\alpha_s^{N+1})$ hard amplitude.

The above procedures are also applicable to the
$O(\alpha_s^{j+1})$ initial-state and final-state distribution
amplitudes $\phi_n^{(j+1)}$ and $\phi_m^{(j+1)}$. We identify the
gluon in the complete set of $O(\alpha_s^{j+1})$ diagrams
$\phi_n^{(j+1)}$, one of whose ends attaches the outer most vertex
$\alpha$ on the $u$ quark line. The other end attaches the vertex
$\beta$ inside the rest of the diagrams. For the first term on the
right-hand side of Eq.~(\ref{dec}), we have a Ward identity
similar to Fig.~5(a). Figure 5(b) then leads to the factorizations
of the initial-state and final-state distribution amplitudes,
\begin{eqnarray}
\phi_{n\parallel}^{(j+1)}&\approx& \sum_{n'=S,T}
\phi_{n'\parallel}^{(1)}\otimes \phi_{n'n}^{(j)}\;,
\label{wep}\\
\phi_{m\parallel}^{(j+1)}&\approx&\sum_{m'=S,T}\phi_{mm'}^{(j)}\otimes
\phi_{m'\parallel}^{(1)}\;,
\label{wi2}
\end{eqnarray}
where the PS and PT structures in Eq.~(\ref{1f}) have been inserted.

We sum Eqs.~(\ref{wic}) and (\ref{wicr}), and subtract the
double-counted diagrams $G^{(N+1)}_{\parallel\;\parallel}$, to
which the replacements are applicable to both the initial-state
and final-state sides as shown in Fig.~7(a). Note that
$G^{(N+1)}_{\parallel\;\parallel}$ does not contain the diagram
in Fig.~7(b), in which the same gluon is identified in
$G^{(N+1)}_{\parallel\cdot}$ and $G^{(N+1)}_{\cdot\parallel}$
simultaneously. This type of diagrams are not double-counted, and
Eq.~(\ref{war2}) does not apply to them. The factorization of the
first type of collinear divergences associated with the identified
radiative gluon from $G_I^{(N+1)}$, $N\ge 1$, is then given by
\begin{eqnarray}
G^{(N+1)}_I&=&G^{(N+1)}_{\parallel\cdot}+G^{(N+1)}_{\cdot \parallel}
-G^{(N+1)}_{\parallel\;\parallel}\;,
\nonumber\\
&\approx&\sum_{n,m=S,T}\left[\phi_{n\parallel}^{(1)}\otimes
G_{nm}^{(N)}\otimes\phi_{m}^{(0)} +\phi_n^{(0)}\otimes
G_{nm}^{(N)}\otimes \phi_{m\parallel}^{(1)}
-\phi_{n\parallel}^{(1)}\otimes G_{nm}^{(N-1)}
\otimes \phi_{m\parallel}^{(1)}\right]\;,
\label{gf}
\end{eqnarray}
where the trivial factorizations, similar to Eq.~(\ref{wi3}), have
been inserted. For the factorization of
$G^{(N+1)}_{\parallel\;\parallel}$, we rely on the Ward identity
in Eq.~(\ref{war2}). Substituting Eqs.~(\ref{gnf1}), (\ref{wep}),
and (\ref{wi2}) into Eq.~(\ref{gf}), a simple algebra gives
\begin{eqnarray}
G^{(N+1)}_I &\approx&\sum_{n,m=S,T}\sum_{i=0}^{N}\sum_{j=0}^{N-i}
\left[\phi_{n\parallel}^{(i+1)}\otimes H_{nm}^{(N-i-j)}\otimes
\phi_{m}^{(j)} +\phi_{n}^{(i)}\otimes H_{nm}^{(N-i-j)}\otimes
\phi_{m\parallel}^{(j+1)}\right.
\nonumber\\
& &\left.-\phi_{n\parallel}^{(i+1)}\otimes H_{nm}^{(N-i-j-1)}\otimes
\phi_{m\parallel}^{(j+1)}\right]\;.
\label{gnf2}
\end{eqnarray}

It is obvious that Eq.~(\ref{gnf2}) is not Lorentz covariant,
since $G^{(N+1)}_I$, $\phi_{n\parallel}^{(i)}$ and
$\phi_{m\parallel}^{(j)}$ include only the first or third term in
Eq.~(\ref{dec}). The Lorentz covariance can be recovered by adding
the other two terms in Eq.~(\ref{dec}) into the tensors for the
identified radiative gluons on both sides of Eq.~(\ref{gnf2}). As
recovering the Lorentz covariance, the second type of collinear
configurations associated with the tensor $-\delta_{\alpha
\perp}\delta_{\beta \perp}$ is taken into account. We then further
add the infrared-finite (self-energy correction) diagrams
mentioned above. At last, the complete collinear factorization of
the diagrams $G^{(N+1)}$ is given by
\begin{eqnarray}
G^{(N+1)}
&=&\sum_{n,m=S,T}\sum_{i=0}^{N}\sum_{j=0}^{N-i}
\left[\phi_{n}^{(i+1)}\otimes H_{nm}^{(N-i-j)}\otimes
\phi_{m}^{(j)} +\phi_{n}^{(i)}\otimes H_{nm}^{(N-i-j)}\otimes
\phi_{m}^{(j+1)}\right.
\nonumber\\
& &\left.-\phi_{n}^{(i+1)}\otimes H_{nm}^{(N-i-j-1)}\otimes
\phi_{m}^{(j+1)}\right]+F^{(N+1)}\;,
\label{gnf3}
\end{eqnarray}
where the $O(\alpha_s^{N+1})$ function $F^{(N+1)}$ contains the
contribution corresponding to the difference between
$\delta_{\alpha +}\delta_{\beta -}$ and $n_{-\alpha}
l_\beta/n_-\cdot l$, except the infrared-finite diagrams.
Equation (\ref{gnf3}) can be simplified into
\begin{eqnarray}
G^{(N+1)}=\sum_{n,m=S,T}\sum_{i=0}^{N+1}\sum_{j=0}^{N+1-i}
\phi_n^{(i)}\otimes H_{nm}^{(N+1-i-j)}\otimes \phi_m^{(j)}\;,
\label{gf1}
\end{eqnarray}
with the $O(\alpha_s^{N+1})$ hard amplitude $H_{nm}^{(N+1)}$ being
defined via
\begin{eqnarray}
F^{(N+1)}=\sum_{n,m=S,T}\phi_n^{(0)}\otimes H_{nm}^{(N+1)} \otimes
\phi_m^{(0)}\;. \label{hn}
\end{eqnarray}
Equation (\ref{gf1}) states that all the two-parton twist-3
collinear divergences in the process $\pi\gamma^*\to\pi$ have been
factorized into the distribution amplitudes in Eq.~(\ref{ld})
order by order.

Before closing this section, we prove that soft divergences do not
exist in the process $\pi\gamma^*\to\pi$ at the two-parton twist-3
level. The $O(\alpha_s)$ soft cancellation has been demonstrated
in Sec.~II. Assume that the $O(\alpha_s^N)$ diagrams $G^{(N)}$ do
not contain any soft divergences, though they contain collinear
ones. They are then dominated by hard dynamics, by collinear
dynamics associated with the initial-state pion, and by collinear
dynamics associated with the final-state pion. Consider the
$O(\alpha_s^{N+1})$ diagrams $G^{(N+1)}$. Similarly, we look for
the radiative gluon radiated from the outer most vertex on the $u$
quark line in the initial state, and adopt the decomposition of
the tensor $g_{\alpha\beta}$ in Eq.~(\ref{dec}). The attachment of
a soft gluon to an off-shell internal line does not introduce an
infrared divergence, since an off-shell propagator behaves at
least like $1/Q$. As the soft gluon attaches a collinear line
along $P_2$, the vertex $\beta$ must be dominated by the minus
component. The gamma matrix $\gamma^\alpha$ is dominated by the
component $\gamma^+$ stated above. Therefore, the replacement in
Eq.~(\ref{rep}) still holds for the first term on the right-hand
side of Eq.~(\ref{dec}). Similarly, the second term on the
right-hand side of Eq.~(\ref{dec}) corresponds to the attachment
of the soft gluon to a collinear line along $P_1$. Again, the
third term on the right-hand side of Eq.~(\ref{dec}) does not
contribute due to the equations of motion.

The above reasoning indicates that the configurations associated
with the identified soft gluon are the same as those associated
with the identified collinear gluon. The procedure for deriving
the collinear factorization applies, and the soft divergences in
the process $\pi\gamma^*\to\pi$, if there are any, can be absorbed
into the initial-state and final-state distribution amplitudes.
According to Eq.~(\ref{gnf}), the distribution amplitudes
$\phi_{n}^{(j)}$ for $j=1$, 2, $\cdots$, $N$ are free of soft
divergences by assumption. We have Eq.~(\ref{wep}) for the
factorization of the $O(\alpha_s^{N+1})$ distribution amplitudes
$\phi^{(N+1)}_{n\parallel}$. Because the soft divergences cancel
among the diagrams for $\phi^{(1)}_{n\parallel}$ shown in Fig.~3
[between Figs.~3(a) and 3(b) and between Figs.~3(c) and 3(d)], and
$\phi_n^{(N)}$ do not contain soft divergences,
$\phi^{(N+1)}_{n\parallel}$ do not either. We then turn to the
$O(\alpha_s^{N+1})$ distribution amplitude associated with the
tensor $-\delta_{\alpha \perp}\delta_{\beta \perp}$. If it
contains soft divergences, the evaluation of these divergences,
arising from an incomplete set of diagrams (the identified soft
gluon does not attach the Wilson line), is not gauge invariant.
This result contradicts the gauge-invariant definition of the
distribution amplitudes in Eq.~(\ref{ld}). Hence, the
$O(\alpha_s^{N+1})$ distribution amplitude must be free of soft
divergences. Extending $N$ to infinity, the absence of the soft
divergences in the distribution amplitudes $\phi_{n,m}$ and in the
diagrams $G$ is proved. We then complete the all-order proof of
the two-parton twist-3 collinear factorization theorem for the
process $\pi\gamma^*\to\pi$.

\section{FACTORIZATION OF $B \to \pi \ell {\bar \nu}$ }

As emphasized in the Introduction, the contributions to the
exclusive semileptonic decay $B\to\pi l\bar\nu$ from the
two-parton twist-3 pion distribution amplitudes $\phi_{S,T}$ are
of the same power as those from the twist-2 one. These
contributions are originally proportional to the ratio $m_0/M_B$.
However, the corresponding convolution integral for the $B\to\pi$
form factor is linearly divergent in collinear factorization
theorem, such that it is proportional to the ratio
$M_B/\bar\Lambda$, if regulated by an effective cut-off $x_c\sim
\bar\Lambda/M_B$. Combining the two ratios $m_0/M_B$ and
$M_B/\bar\Lambda$, the two-parton twist-3 contributions are in
fact not down by a power of $1/M_B$:
\begin{eqnarray}
\frac{m_0}{M_B}\int_{x_c}^1 dx_2H_S(x_2)\phi_S(x_2)\sim
\frac{m_0}{M_B}\int_{x_c}^1 dx_2\frac{1}{x_2^2}\sim
O\left(\frac{m_0}{\bar\Lambda}\right)\;, \label{li}
\end{eqnarray}
for the asymptotic functional form $\phi_S\sim 1$ \cite{PB1} and
$H_S\sim 1/x_2^2$ in Eq.~(\ref{bpb}) below. The presence of the
linear divergences modifies the naive power counting rules, and
the two-parton twist-3 contributions become leading-power in the
$B$ meson transition form factors \cite{TLS}. This is our
motivation to prove the corresponding factorization theorem. Note
that the two-parton twist-3 contributions are
next-to-leading-power in the pion form factor.

The momentum $P_1$ of the $B$ meson and the momentum $P_2$ of the
outgoing pion are parameterized as
\begin{eqnarray}
P_1=\frac{M_B}{\sqrt{2}}(1,1,{\bf 0}_T)\;,\;\;\;
P_2=\frac{M_B}{\sqrt{2}}(0,\eta,{\bf 0}_T)\;,
\label{bmpp}
\end{eqnarray}
where $\eta$ denotes the energy fraction carried by the pion.
Consider the kinematic region with small $q^2$, $q=P_1-P_2$ being
the lepton pair momentum, {\it i.e.}, with large $\eta$, where
PQCD is applicable. Let the light spectator $\bar d$ quark in the
$B$ meson (pion) carry the momentum $k_1$ ($k_2=x_2P_2$), and the
$b$ quark carry the momentum $P_1-k_1$. We have the equations of
motion,
\begin{eqnarray}
(\not\! P_1-\not k_1-m_b)b(P_1-k_1)=0\;,\;\;\;\;
{\bar d}(k_1)\not\! k_1=0\;,
\label{eqb}
\end{eqnarray}
where the $\bar d$ quark mass $m_d$ has been neglected, and those
for the valence quarks in the pion.

\subsection{$O(\alpha_s)$ Factorization}

We start with the $O(\alpha_s)$ collinear factorization of the
final-state distribution amplitudes. The lowest-order diagrams for
the $B\to\pi l\bar\nu$ decay are the same as in Fig.~1, but with the
upper quark line in the initial state representing a $b$ quark and with
the symbol $\times$ representing a weak decay vertex. Figure~1(a) gives
the amplitude,
\begin{eqnarray}
G^{(0)}(x_1,x_2)&=&\frac{-g^2C_F}{2} \frac{u({\bar
x}_2P_2)\gamma^\nu(\not\! P_2-\not k_1)\gamma_\mu b(P_1-k_1){\bar
d}(k_1)\gamma_\nu {\bar d}(x_2P_2)} {(P_2-k_1)^2(k_1-x_2P_2)^2}\;,
\label{b1a}
\end{eqnarray}
where the higher-power term $\not\! k_1$ in the numerator has been
dropped. Inserting the Fierz identity in Eq.~(\ref{1f}) into the
above expression, we obtain the trivial factorization formula,
\begin{eqnarray}
G^{(0)}(x_1,x_2)= \int d\xi_2H^{(0)}_{S}(x_1,\xi_2)
\phi_S^{(0)}(x_2,\xi_2)\;,
\end{eqnarray}
with the lowest-order hard amplitude and distribution amplitude,
\begin{eqnarray}
H_{S}^{(0)}(x_1,\xi_2) &= & \frac{-g^2C_F}{2}m_0 \frac{
tr[\gamma^\nu(\not\! P_2-\not\! k_1)\gamma_\mu b(P_1-k_1){\bar
d}(k_1)\gamma_\nu\gamma^5]} {(P_2-k_1)^2(k_1-x_2P_2+l)^2}
\nonumber \\
&=&\frac{-g^2C_F}{2\eta^2M_B^4}m_0 \frac{tr[\gamma^\nu\not\!
P_2\gamma_\mu b(P_1-k_1){\bar
d}(k_1)\gamma_\nu\gamma^5]}{x_1^2\xi_2}\;,
\nonumber\\
\phi_S^{(0)}(x_2,\xi_2)&=&\frac{1}{4m_0}u({\bar x}_2P_2)\gamma_5
{\bar d}(x_2P_2)\delta(\xi_2-x_2)\;. \label{b1aS}
\end{eqnarray}
The PT structure does not contribute because of $\gamma^\nu(\not\!
n_+\not\! n_--1)\gamma_5\gamma_\nu=0$. It is observed that
$H_S^{(0)}$ depends only on the single plus component of $k_1$
through $k_1\cdot P_2$, which defines the momentum fraction
$x_1=k_1^+/P_1^+$ carried by the $\bar d$ quark in the $B$ meson.

Consider $O(\alpha_s)$ corrections to Fig.~1(a), which are
displayed in Fig.~4 with the initial and final states being
flipped. Here we summarize only the results of their
factorization, and refer the details to Appendix C. The
factorization of the two-particle reducible diagrams in
Figs.~4(a)-4(c) is straightforward. After inserting the Fierz
identity, the loop integrals associated with Figs.~4(a), 4(b), and
4(c) are written as
\begin{eqnarray}
I^{(a),(b),(c)}&\approx& \int d\xi_2
H_S^{(0)}(x_1,\xi_2)\phi^{(1)}_{Sa,Sb,Sc}(x_2,\xi_2)\;,
\label{4a2}
\end{eqnarray}
where the PS collinear pieces are given by
\begin{eqnarray}
\phi_{Sa}^{(1)}(x_2,\xi_2)\!\!\!&=&\!\!\!\frac{-ig^2C_F}{8m_0}
\!\!\int\!\frac{d^4l}{(2\pi)^4}u({\bar x}_2P_2)\gamma_\beta
\frac{{\bar x}_2\not\! P_2+\not l}{({\bar x}_2P_2+l)^2} \gamma^\beta
\frac{1}{\bar{x}_2\not\! P_2} \gamma_5{\bar d}(x_2P_2)\frac{1}{l^2}
\delta(\xi_2-x_2)\;,
\label{p4a}\\
\phi_{Sb}^{(1)}(x_2,\xi_2)\!\!\!&=&\!\!\!\frac{ig^2C_F}{4m_0}
\!\!\int\!\frac{d^4l}{(2\pi)^4}u({\bar x}_2P_2)\gamma_\beta
\frac{{\bar x}_2\not\! P_2+\not l}{({\bar x}_2P_2+l)^2}\gamma_5
\frac{x_2\not\! P_2-\not l}{(x_2P_2-l)^2}\gamma^\beta
{\bar d}(x_2P_2)\frac{1}{l^2}
\delta\left(\xi_2-x_2+\frac{l^-}{P_2^-}\right)\;,
\label{p4b} \\
\phi_{Sc}^{(1)}(x_2,\xi_2)\!\!\!&=&\!\!\!\frac{-ig^2C_F}{8m_0}
\!\!\int\!\frac{d^4l}{(2\pi)^4}u({\bar x}_2P_2)
\gamma_5\frac{1}{x_2\not\! P_2} \gamma_\beta\frac{x_2\not\! P_2-\not
l}{(x_2P_2-l)^2}\gamma^\beta {\bar d}(x_2P_2)\frac{1}{l^2}
\delta(\xi_2-x_2)\;. \label{p4c}
\end{eqnarray}
The momentum fraction $\xi_2$ in Fig.~4(b) has been modified into
$\xi_2=x_2-l^-/P_2^-$ by the collinear gluon exchange.

The collinear factorization of Figs.~4(d)-4(g) is summarized as
\begin{eqnarray}
I^{(d),(f)} &\approx & \int d\xi_2
H_{S}^{(0)}(x_1,\xi_2)\phi_{Sd,Sf}^{(1)}(x_2,\xi_2)\;,
\nonumber\\
I^{(e)}+I^{(g)} &\approx & \int d\xi_2
H_{S}^{(0)}(x_1,\xi_2)\phi_{Seg}^{(1)}(x_2,\xi_2)\;,
\label{4f2}
\end{eqnarray}
with the PS pieces,
\begin{eqnarray}
\phi_{Sd}^{(1)}(x_2,\xi_2)\!\!\!&=&\!\!\!\frac{-ig^2}{2m_0C_F}
\!\int\frac{d^4l}{(2\pi)^4}u({\bar x}_2P_2)\gamma^\beta
\frac{{\bar x}_2\not\! P_2 +\not l}{({\bar x}_2P_2+l)^2}
\gamma_5 {\bar d}(x_2P_2)\frac{1}{l^2}
\frac{n_{+\beta}}{n_+\cdot l}
\nonumber\\
& &\times\left[\delta(\xi_2-x_2)-
\delta\left(\xi_2-x_2+\frac{l^-}{P_2^-}\right)\right]\;,
\label{pb4d}\\
\phi_{Seg}^{(1)}(x_2,\xi_2)
\!\!\!&=&\!\!\!\frac{ig^2}{8m_0N_c}
\!\int\frac{d^4l}{(2\pi)^4}u({\bar x}_2P_2)\gamma^\beta
\frac{{\bar x}_2\not\! P_2+\not l}{({\bar x}_2P_2+l)^2}\gamma_5
{\bar d}(x_2P_2)\frac{1}{l^2}
\frac{n_{+\beta}}{n_+\cdot l}\delta(\xi_2-x_2)\;,
\label{pb4eg}\\
\phi_{Sf}^{(1)}(x_2,\xi_2)
\!\!\!&=&\!\!\!\frac{-ig^2}{8m_0N_c}
\!\int\frac{d^4l}{(2\pi)^4}u({\bar x}_2P_2)\gamma^\beta
\frac{{\bar x}_2\not\! P_2+\not l}{({\bar x}_2P_2+l)^2}\gamma_5
{\bar d}(x_2P_2)\frac{1}{l^2}
\frac{n_{+\beta}}{n_+\cdot l}
\delta\left(\xi_2-x_2+\frac{l^-}{P_2^-}\right)\;.
\label{pb4f}
\end{eqnarray}
The contribution from Fig.~4(d) has been split into two terms as a
consequence of the Ward identity. The first and second terms
correspond to the cases without and with the loop momentum $l$
flowing through the hard gluon, respectively. The Feynman rule
$n_{+\beta}/n_+\cdot l$ in the collinear divergent pieces, coming
from the eikonal approximation, can be represented by a Wilson
line in the direction $n_+$. Note that each of the hard amplitudes
from Figs.~4(e) and 4(g) contains a residual dependence on $l$. It
is their sum that does not depend on $l$, and appears in the
desired factorization form shown in Eq.~(\ref{4f2}) (see Appendix
C).

Equations (\ref{pb4d})-(\ref{pb4f})
carry different color factors due to different color flows in
Figs.~4(d)-4(g). Their sum leads to the factorization in the
color space:
\begin{eqnarray}
\sum_{i=(d)}^{(g)}I^{i}&\approx&\int d\xi_2
H_S^{(0)}(x_1,\xi_2)\phi_{Su}^{(1)}(x_2,\xi_2)\;,
\label{4bdg}
\end{eqnarray}
where the PS collinear piece,
\begin{eqnarray}
\phi_{Su}^{(1)}(x_2,\xi_2)&=&\frac{-ig^2C_F}{4m_0}
\int\frac{d^4l}{(2\pi)^4}u({\bar x}_2P_2)\gamma^\beta
\frac{{\bar x}_2\not\! P_2 +\not l}{({\bar x}_2P_2 +l)^2}\gamma_5
{\bar d}(x_2P_2) \frac{1}{l^2}
\frac{n_{+\beta}}{n_+\cdot l}
\nonumber\\
& &\times \left[\delta(\xi_2-x_2)-
\delta\left(\xi_2-x_2+\frac{l^-}{P_2^-}\right)\right]\;,
\label{4nu}
\end{eqnarray}
is associated with the collinear gluon emitted from the $u$ quark.
The appropriate color factor $C_F$ indicates the factorization between
the distribution amplitudes and the hard amplitude in color space.

The collinear factorization of Figs.~4(h)-4(k), derived in a similar
way, is written as
\begin{eqnarray}
\sum_{i=(h)}^{(k)}I^{i}&\approx&\int d\xi_2
H_S^{(0)}(x_1,\xi_2)\phi_{S\bar d}^{(1)}(x_2,\xi_2)\;,
\label{4bhk}
\end{eqnarray}
where the PS piece,
\begin{eqnarray}
\phi_{S\bar d}^{(1)}(x_2,\xi_2)&=&\frac{ig^2C_F}{4m_0}
\int\frac{d^4l}{(2\pi)^4}u({\bar x}_2P_2)
\gamma_5\frac{x_2\not\! P_2 -\not l}{(x_2P_2 -l)^2}
\gamma^\beta {\bar d}(x_2P_2)\frac{1}{l^2}
\frac{n_{+\beta}}{n_+\cdot l}
\nonumber\\
& &\times \left[\delta(\xi_2-x_2)-
\delta\left(\xi_2-x_2+\frac{l^-}{P_2^-}\right)\right]\;,
\label{4nd}
\end{eqnarray}
is associated with the collinear gluon emitted from the $\bar d$ quark.
Note that Fig.~4(k) also contributes a collinear divergence, and we
have to combine the results from Figs.~4(j) and 4(k), so that the hard
amplitude does not depend on the loop momentum $l$. At last, the sum of
Eqs.~(\ref{4a2}), (\ref{4bdg}) and (\ref{4bhk}) gives
\begin{eqnarray}
\sum_{i=(a)}^{(k)}I^{i}\approx\int d\xi_2
H_S^{(0)}(x_1,\xi_2)\phi_S^{(1)}(x_2,\xi_2)\;,
\end{eqnarray}
where the PS collinear piece $\phi_S^{(1)}$ is defined by the
$O(\alpha_s)$ term of the complex
conjugate of Eq.~(\ref{ld}), consistent with the universality.

The amplitude corresponding to Fig.~1(b) is written as
\begin{eqnarray}
G^{(0)}(x_1,x_2)&=&\frac{-g^2 C_F}{2}
\frac{u({\bar x}_2P_2)\gamma_\mu(\not\! P_1-x_2\not\! P_2+m_b)
\gamma^\nu b(P_1-k_1){\bar d}(k_1)\gamma_\nu {\bar d}(x_2P_2)}
{[(P_1-x_2P_2)^2-m_b^2](k_1-x_2P_2)^2}\;,
\label{b1b}
\end{eqnarray}
which, after inserting the Fierz identity, leads to the trivial
factorization formula,
\begin{eqnarray}
G^{(0)}= \sum_{m=S,T}\int d\xi_2
H^{(0)}_{m}(x_1,\xi_2)\phi_m^{(0)}(x_2,\xi_2)\;.
\end{eqnarray}
The lowest-order hard amplitudes and distribution amplitudes
of the PS and PT structures are written as,
\begin{eqnarray}
H_{S}^{(0)}(x_1,\xi_2)&=&\frac{-g^2 C_F}{2}m_0\frac{tr[\gamma_\mu
(\not\! P_1-\xi_2\not\! P_2+m_b)\gamma^\nu b(P_1-k_1){\bar d}(k_1)
\gamma_\nu\gamma_5]}{[(P_1-\xi_2P_2)^2-m_b^2](k_1-\xi_2P_2)^2}\;,
\nonumber\\
&=& \frac{-g^2 C_F}{2\eta^2M_B^4}m_0\frac{tr[\gamma_\mu (\not\!
P_1-\xi_2\not\! P_2+m_b)\gamma^\nu b(P_1-k_1){\bar d}(k_1)
\gamma_\nu\gamma_5]}{x_1\xi_2^2}\;,
\nonumber\\
H_{T}^{(0)}(x_1,\xi_2)&=&\frac{-g^2 C_F}{2}m_0\frac{tr[\gamma_\mu
(\not\! P_1-\xi_2\not\! P_2+m_b)\gamma^\nu b(P_1-k_1){\bar d}(k_1)
\gamma_\nu(\not\! n_+\not\! n_--1)\gamma_5]}
{[(P_1-\xi_2P_2)^2-m_b^2](k_1-\xi_2P_2)^2}\;,
\nonumber\\
\phi_T^{(0)}(x_2,\xi_2)&=&\frac{1}{4m_0}u({\bar x}_2P_2)
\gamma_5(\not\! n_+\not\! n_--1){\bar d}(x_2P_2) \delta(\xi_2-x_2)\;.
\label{bpb}
\end{eqnarray}
The lowest-order PS distribution amplitude $\phi_S^{(0)}(x_2,\xi_2)$ is
the same as in Eq.~(\ref{b1aS}).

Below we discuss the collinear divergences in the $O(\alpha_s)$
corrections to Fig.~1(b) in the covariant gauge, which are
displayed in Fig.~2 with the initial states and the final states
being flipped. The details are referred to Appendix C. Figures 2(a)-2(c)
are factorized straightforwardly, leading to
\begin{eqnarray}
I^{(a),(b),(c)}&\approx& \sum_{m=S,T}\int d\xi_2
H_m^{(0)}(x_1,\xi_2)\phi^{(1)}_{ma,mb,mc}(x_2,\xi_2)\;.
\label{b2b}
\end{eqnarray}
The PS collinear divergent functions
$\phi^{(1)}_{Sa,Sb,Sc}\;(x_2,\xi_2)$ are the same as those shown in
Eqs.~(\ref{p4a}) - (\ref{p4c}), respectively. The PT pieces have the
similar expressions with $\gamma_5$ being replaced by
$\gamma_5\;(\not\! n_+\not\! n_--1)$.

For the irreducible diagrams Figs.~2(d)-2(g), a summation of their
contributions is necessary for obtaining the desired collinear
factorization. Figure~2(d) is split into two terms after applying
the Ward identity, similar to Eq.~(\ref{pb4d}). The sum of the
first term and Fig.~2(e) gives the correct color factor, which
corresponds to the separate color flows between the distribution
amplitudes and the hard amplitude. However, the hard amplitude
still depends on the loop momentum $l$, which is not yet in the
desired form. We must further add Fig.~2(g) to this sum in order
to obtain the hard amplitude without the $l$ dependence. The sum
of the second term from Fig.~2(d) and Fig.~2(f) also gives the
correct color factor, and the result represents the case with $l$
flowing through the hard amplitude. Hence, we arrive at, from the
combination of Figs.~2(d)-2(g),
\begin{eqnarray}
\sum_{i=(d)}^{(g)}I^{i}&\approx& \sum_{m=S,T}\int d\xi_2
H_m^{(0)}(x_1,\xi_2)\phi_{mu}^{(1)}(x_2,\xi_2)\;.
\label{2bdg}
\end{eqnarray}
The collinear factorization of Fig.~2(h)-2(k), derived in
a similar way, is written as
\begin{eqnarray}
\sum_{i=(h)}^{(k)}I^{i}&\approx& \sum_{m=S,T}\int d\xi_2
H_m^{(0)}(x_1,\xi_2)\phi_{m{\bar d}}^{(1)}(x_2,\xi_2)\;,
\label{2bhk}
\end{eqnarray}
with the PS pieces $\phi_{Su}^{(1)}$ and $\phi_{S\bar d}^{(1)}$
shown in Eqs.~(\ref{4nu}) and (\ref{4nd}), respectively. At last, the
sum of Eqs.~(\ref{b2b}), (\ref{2bdg}) and (\ref{2bhk}) gives
\begin{eqnarray}
\sum_{i=(a)}^{(k)}I^{i}\approx
\sum_{m=S,T}\int d\xi_2H_m^{(0)}(x_1,\xi_2)\phi_m^{(1)}(x_2,\xi_2)\;,
\label{phi2}
\end{eqnarray}
where the collinear divergent functions $\phi_m^{(1)}(x_2,\xi_2)$
are defined by the $O(\alpha_s)$ terms of
the complex conjugate of Eq.~(\ref{ld}).

The $O(\alpha_s)$ factorization of the soft divergences from the
decay $B\to\pi l\bar\nu$ has been performed in \cite{L1}, which
results in two $B$ meson distribution amplitudes $\phi_+^{(1)}$
and $\phi_-^{(1)}$ \cite{BF,GN} arising from the insertion of the
fourth and fifth terms of the Fierz identity on the initial-state
side:
\begin{eqnarray}
I_{ij}I_{lk}&=&\frac{1}{4}I_{ik}I_{lj}
+\frac{1}{4}(\gamma_\alpha)_{ik}(\gamma^\alpha)_{lj}
+\frac{1}{4}\left[\frac{1}{\sqrt{2}}\gamma_5(\not
v-1)\right]_{ik}\left[\frac{1}{\sqrt{2}}(\not
v-1)\gamma_5\right]_{lj}
\nonumber\\
& &+\frac{1}{4}\left[\frac{1}{\sqrt{2}}\gamma_5\not\! n_+(\not
v+1)\right]_{ik}\left[\frac{1}{\sqrt{2}}(\not v+1)\not\!
n_-\gamma_5\right]_{lj}
\nonumber\\
& &+\frac{1}{4}\left[\frac{1}{\sqrt{2}}\gamma_5\not\! n_-(\not
v+1)\right]_{ik}\left[\frac{1}{\sqrt{2}}(\not v+1)\not\!
n_+\gamma_5\right]_{lj}\;. \label{2f}
\end{eqnarray}
The Wilson line on the light cone can be constructed, only if the
hard scale for exclusive $B$ meson decays is of
$O(\sqrt{\bar\Lambda M_B})$. This Wilson line is crucial for the
gauge-invariant definitions of the distribution amplitudes as
nonlocal matrix elements.

Following the similar procedures in Sec.~II, we derive the
$O(\alpha_s)$ factorization of the decay $B\to \pi l\bar \nu$,
\begin{eqnarray}
G^{(1)}=\sum_{n=+,-}\phi^{(1)}_n\otimes H^{(0)}_n
+\sum_{m=S,T}H^{(0)}_m\otimes\phi^{(1)}_m+H^{(1)}\;,
\end{eqnarray}
where the hard amplitude $H^{(0)}$ receives the contributions from both
Fig.~1(a) and 1(b). Consequently, the factorization formula for the
two-parton twist-3 contributions is written, up to $O(\alpha_s)$, as
\begin{eqnarray}
G^{(0)}+G^{(1)}=\sum_{n=+,- \atopwithdelims . .m=S,T}
(\phi^{(0)}_n+\phi^{(1)}_n)
\otimes(H_{nm}^{(0)}+H^{(1)}_{nm})\otimes
(\phi^{(0)}_m+\phi^{(1)}_m)\;.
\end{eqnarray}
The definitions for the hard amplitudes $H_{nm}^{(1)}$ and
$H_{nm}^{(0)}$ are similar to those in Eq.~(\ref{wi3}). For
example, the explicit expression of $H_{+S}^{(0)}$ is given by
\begin{eqnarray}
H_{+S}^{(0)}(\xi_1,\xi_2) &= & \frac{-g^2C_F}{2\sqrt{2}}m_0
\Bigg\{\frac{ tr[\gamma^\nu(\not\! P_2-\not k_1)\gamma_\mu (\not
\!P_1+M_B)\not\! n_-\gamma_5\gamma_\nu\gamma_5]}
{(P_2-k_1)^2(k_1-x_2P_2+l)^2}
\nonumber\\
& &+ \frac{tr[\gamma_\mu (\not\! P_1-\xi_2\not\! P_2+m_b)\gamma^\nu
(\not\! P_1+M_B)\not\! n_-\gamma_5
\gamma_\nu\gamma_5]}{[(P_1-\xi_2P_2)^2-m_b^2](k_1-\xi_2P_2)^2}\Bigg\}\;.
\label{exb}
\end{eqnarray}

\subsection{All-order Factorization}

The all-order proof of the two-parton twist-3 factorization
theorem for the process $\pi\gamma^*\to\pi$ in Sec.~III can be
generalized to the $B\to\pi l\bar\nu$ decay with minor
modifications. Here we highlight only the different points of the
proof. In the case of $B$ meson decays there is no collinear
divergence associated with the initial state, since the $b$ quark
is massive, and the light spectator $\bar d$ quark is soft
\cite{L1}. Hence, the $B$ meson side is dominated by the soft
divergence. The collinear configurations associated with the
final-state pion are the same as in the process
$\pi\gamma^*\to\pi$ discussed in Sec.~II. The important infrared
divergences are then classified into the soft type with small loop
momentum $l$ and the collinear type with $l$ parallel to $P_2$. We
shall compare the factorizations of the soft divergences into the
initial state and of the collinear divergences into the final
state.

Identify the soft gluon emitted from the outer most vertex
$\alpha$ on the $b$ quark line in the $O(\alpha_s^{N+1})$ diagrams
$G^{(N+1)}$. Let $\beta$ denote the attachments of the other end
of the identified gluon inside the diagrams. The attachment of the
soft gluon to collinear lines along $P_2$ and to soft lines gives
soft divergences. The soft lines include the $b$ quark line, the
spectator $\bar d$ quark line, and soft internal lines. The
attachment to hard lines off-shell by
$O((P_2-k_1)^2)\sim O(\bar\Lambda M_B)$ also gives soft divergences
\cite{L1}. When the identified gluon attaches the collinear lines
along $P_2$ and the hard lines along $P_2-k_1$, the vertex $\beta$
inside $G^{(N)}$ is mainly minus, and the vertex $\alpha$ on the
$b$ quark line is mainly plus. Therefore, the replacement in
Eq.~(\ref{rep}) still works. Certainly, Eq.~(\ref{rep}) changes the
soft divergences from the attachment of the identified gluon to the
soft lines. However, these soft lines appear only on the $B$ meson
side, and will be compensated, when recovering the Lorentz covariance,
a case similar to the factorization of the second type of collinear
configurations in the process $\pi\gamma^*\to\pi$.

For the identified soft gluon emitted by the spectator $\bar d$
quark, the first term $\delta_{\alpha +}\delta_{\beta -}$ on the
right-hand side of Eq.~(\ref{dec}) corresponds to the
configuration, where the identified gluon attaches all types of
lines except the spectator $\bar d$ quark line. This
configuration is the same as that associated with the soft
gluon emission from the $b$ quark stated above. That is,
Eq.~(\ref{rep}) holds. The second term
$-\delta_{\alpha \perp}\delta_{\beta \perp}$ corresponds to the
configuration with the identified gluon attaching the soft lines.
The third term $\delta_{\alpha
-}\delta_{\beta +}$ does not contribute because of the equations
of motion.

Since the collinear configurations for the final-state side are
the same as those in the process $\pi\gamma^*\to \pi$,
Eq.~(\ref{rep1}) holds. Based on the above discussion, the symbol
$\parallel$ in $B\to\pi l\bar\nu$ has the same meaning as in
$\pi\gamma^*\to \pi$, and the treatment of $G^{(N+1)}_{I}$ follows
the procedure in Sec.~III [through Figs.~5(a) and 5(b)]. The above
reasoning also applies to the factorization of the
$O(\alpha_s^{(j+1)})$ initial-state distribution amplitudes
$\phi_{n\parallel}^{(j+1)}$. In summary, we obtain the similar
factorizations but with the subscript $n=+,-$ for the initial
state

Following the same induction procedure, we obtain the
factorization of the full diagrams $G^{(N+1)}$,
\begin{eqnarray}
G^{(N+1)}=\sum_{n=+,- \atopwithdelims . .m=S,T}
\sum_{i=0}^{N+1}\sum_{j=0}^{N+1-i}
\phi_n^{(i)}\otimes H_{nm}^{(N+1-i-j)}\otimes \phi_m^{(j)}\;,
\label{gf2}
\end{eqnarray}
where the $O(\alpha_s^{N+1})$ hard amplitude $H_{nm}^{(N+1)}$ is
infrared finite. Equation (\ref{gf2}) indicates that all the soft
and collinear divergences in the semileptonic decay $B\to\pi
l\bar\nu$ can be factorized into the distribution amplitudes
$\phi_n^{(i)}$ and $\phi_m^{(j)}$ at the parton level order by
order, and that the proof of the corresponding two-parton twist-3
factorization theorem is completed. These parton-level
distribution amplitudes serve as the infrared regulators for the
derivation of the hard amplitudes from the parton-level diagrams.
We then compute the $B\to\pi$ transition form factor by
convoluting the hard amplitudes with the meson distribution
amplitudes, in which the quark states are replaced by the physical
$B$ meson and pion states. Both the twist-2 and two-parton twist-3
contributions to the $B\to\pi$ form factors $F_+(q^2)$ and
$F_0(q^2)$ in the standard definition have been evaluated in
\cite{TLS}.
It has been observed that the latter are of the same order of
magnitude as the former, consistent with the argument that the
two-parton twist-3 contributions are not power-suppressed and
chirally enhanced. The light-cone sum rules also give
approximately equal weights to the twist-2 and two-parton twist-3
contributions to $F_+$ \cite{KR2}.


\section{CONCLUSION}

In this paper we have investigated the infrared divergences in the
process $\pi\gamma^*\to\pi$ at the two-parton twist-3 level. We
summarize our observations below. There are no soft divergences
associated with the pion, since they cancel among diagrams. The
absence of the soft divergences is related to the fact that a soft
gluon, being huge in space-time, does not resolve the color
structure of the color-singlet pion. In the collinear region with
the loop momentum parallel to the pion momentum, we have shown
that the delicate summation of many diagrams leads to the
$O(\alpha_s)$ factorization in the momentum, spin and color
spaces. We have presented an all-order proof of the two-parton
twist-3 factorization theorem for the process $\pi\gamma^*\to\pi$
by means of the Ward identity. This proof can also accommodate the
twist-2 factorization theorem presented in \cite{L1} and the
twist-4 one simply by considering the corresponding structures in
the Fierz transformation in Eq.~(\ref{1f}).

The idea of the proof is to decompose the tensor $g_{\alpha\beta}$
for the identified collinear gluon into the longitudinal and
transverse pieces shown in Eq.~(\ref{dec}). The longitudinal
(transverse) piece corresponds to the configuration without (with)
the attachment of the identified gluon to a line along the
external momentum. The former configuration can be factorized
using the Ward identity as hinted by the replacement in
Eq.~(\ref{rep}). The factorization of the latter configuration can
be included by demanding the Lorentz covariance of the
factorization. We emphasize again that the parton-level
distribution amplitudes serve as the infrared regulators for the
derivation of the hard amplitudes from the parton-level diagrams,
similar to the effective diagrams drawn in SCET \cite{bfl,bfps}.
The hard amplitudes are then derived from the ``matching
procedure". Based on the perturbative construction of the
distribution amplitudes, we have derived their two-parton twist-3
definitions as nonlocal matrix elements, where the path-ordered
Wilson line appears as a consequence of the Ward identity.  Note
that our technique is simple compared to that based on the
''$\Delta$-forest" prescription in \cite{DM}, and explicitly gauge
invariant compared to that performed in the axial gauge \cite{BL}.

We have generalized the proof to the more complicated semileptonic
decay $B\to\pi l\bar\nu$. The collinear factorization for the
final-state pion is the same as in the process
$\pi\gamma^*\to\pi$. The identical collinear structures in both
processes justify the concept of universality of
hadron distribution amplitudes in PQCD. The factorization of the
soft divergences for the initial-state $B$ meson has been
discussed in \cite{L1}. The conceptual differences are summarized
as follows. The attachments of a soft gluon to the hard lines
off-shell by $O(\bar\Lambda M_B)$ lead to soft divergences. These
divergences, like those from the attachments of a collinear gluon
to the hard lines in the process $\pi\gamma^*\to\pi$, are crucial
for constructing the Wilson line, that guarantees the gauge
invariance of the $B$ meson distribution amplitudes. This explains 
why the characteristic scale of exclusive $B$ meson decays, if
factorizable, must be of $O(\bar\Lambda M_B)$. The decomposition of 
the tensor $g_{\alpha\beta}$ for the identified soft gluon in 
Eq.~(\ref{dec}) and the replacement in Eq.~(\ref{rep}) still work. 
The procedures of the proof then follow those for the pion form factor.


For a practical application to the $B\to\pi l\bar\nu$ decay, the
parton transverse momenta $k_T$ must be taken into account in
order to smear the end-point singularities in the hard amplitudes
\cite{TLS,LY1}. This observation implies the necessity of proving
$k_T$ factorization theorem \cite{BS,LS}. The proof of $k_T$
factorization theorem is basically the same as proposed in this
paper: we simply retain the dependence on the loop transverse
momenta in hard amplitudes \cite{NL}. The relative importance of
the twist-2 and two-parton twist-3 contributions to the $B\to\pi$
transition form factor has been investigated in \cite{TLS}, which
confirms our motivation to prove the two-parton twist-3
factorization theorem: the latter contributions are not
power-suppressed and chirally enhanced. In a future work the proof
will be generalized to nonleptonic $B$ meson decays, such as $B\to
K\pi$ and $\pi\pi$ \cite{KLS,LUY}. The corresponding factorization
theorem is more complicated, since nonleptonic decays involve
three characteristic scales: the $W$ boson mass $M_W$, $M_B$, and
the factorization scale of $O(\bar\Lambda)$, such as the parton
transverse momenta $k_T$ \cite{CL,YL}.

At last, we compare our construction of collinear factorization
theorem in perturbation theory with that in SCET. In the former one
starts with Feynman diagrams in full QCD. Look for the leading
region of the loop momentum defined by Eq.~(\ref{sog2}), in which
one makes the power counting of the Feynman diagrams. It can be
found that the approximate loop integral in the leading region is
represented by the diagram in Fig.~3(e), which leads to the
definition of a distribution amplitude. In SCET one first
constructs the various effective degrees of freedom describing
infrared dynamics and the effective interactions, and defines
their powers. Select a specific effective operator, such as those
nonlocal operators in Eq.~(\ref{pt}). Draw the diagrams based on
the effective theory, and then make the power counting. It can be
shown that the diagram in Fig.~3(e) scales like the selected
operator, and builds up the distribution amplitude. It is not
necessary to analyze the infrared divergences in diagrams at this
stage. That is, one arrives at Fig.~3(e) through approximating
loop integrals in the full theory in PQCD, but does at the
operator and Lagrange level in SCET. Despite of the
different reasonings for deriving a collinear factorization
formula, the calculation of the Wilson coefficients is the same.
As calculating the Wilson coefficients associated with the
effective operators from the matching procedure in SCET, the
infrared divergences need to be analyzed, and their cancellation
between the full theory and the effective theory must be
demonstrated explicitly. This procedure is in fact the same as the
derivation of the hard amplitudes (Wilion coefficients) in PQCD,
where the subtraction of the distribution amplitudes (the
effective theory) from the parton-level diagrams (the full theory) is
done. Therefore, it is legitimate to claim that the constructions of
collinear factorization theorem are equivalent between PQCD
and SCET \cite{Li03}.

\vskip 0.5cm

We thank S. Brodsky, A.I. Sanda, and G. Sterman for useful
discussions. This work was supported in part by the National
Science Council of R.O.C. under the Grant No.
NSC-92-2112-M-001-030, by National Center for Theoretical Sciences
of R.O.C., and by Grant-in Aid for Special Project Research
(Physics of CP Violation) from the Ministry of Education, Science
and Culture, Japan.

\appendix

\section{$O(\alpha_s)$ CORRECTIONS FROM FIG.~2}

In this Appendix we present the details of the collinear
factorization of Fig.~2 associated with the initial state. The
loop integral of Fig.~2(b) is given by
\begin{eqnarray}
I^{(b)}&=& -\frac{1}{2}eg^4 C_F^2\int \frac{d^4l}{(2\pi)^4} {\bar
d}(x_1P_1)\gamma_\beta \frac{x_1\not\! P_1-\not l}{(x_1P_1-l)^2}
\gamma_\alpha {\bar d}(x_2 P_2)u({\bar x}_2 P_2)
\nonumber\\
& &\times \gamma^\alpha\frac{\not\! P_2-x_1\not\! P_1+\not
l}{(P_2-x_1P_1+l)^2} \gamma_\mu \frac{{\bar x}_1\not\! P_1 +\not
l}{({\bar x}_1 P_1 +l)^2} \gamma^\beta u({\bar
x}_1P_1)\frac{1}{l^2(x_2P_2-x_1P_1+l)^2}\;. \label{2b}
\end{eqnarray}
Inserting the Fierz identity in Eq.~(\ref{1f}), we obtain
\begin{eqnarray}
I^{(b)}&\approx&\frac{ig^2 C_F}{4m_0}
\int\frac{d^4l}{(2\pi)^4}{\bar d}(x_1P_1)\gamma_\beta
\frac{x_1\not\! P_1-\not l}{(x_1P_1-l)^2}\gamma_5 \frac{{\bar
x}_1\not\! P_1 +\not l}{({\bar x}_1 P_1 +l)^2}\gamma^\beta u({\bar
x}_1P_1)\frac{1}{l^2}
\nonumber \\
& & \;\;\times\frac{i}{2}eg^2C_Fm_0\frac{{\rm tr}[\gamma_\alpha
{\bar d}(x_2 P_2)u({\bar x}_2 P_2) \gamma^\alpha(\not\! P_2-x_1\not\!
P_1+\not l)\gamma_\mu
\gamma_5]}{(P_2-x_1P_1+l)^2(x_2P_2-x_1P_1+l)^2}
\nonumber\\
&&+ \frac{ig^2 C_F}{4m_0} \int\frac{d^4l}{(2\pi)^4}{\bar
d}(x_1P_1)\gamma_\beta \frac{x_1\not\! P_1-\not
l}{(x_1P_1-l)^2}\gamma_5 (\not n_+\!\!\not n_-\!\!-1)\frac{{\bar
x}_1\not\! P_1 +\not l}{({\bar x}_1 P_1 +l)^2}\gamma^\beta u({\bar
x}_1P_1)\frac{1}{l^2}
\nonumber \\
& & \;\;\times\frac{i}{2}eg^2C_Fm_0 \frac{{\rm tr}[\gamma_\alpha
{\bar d}(x_2 P_2)u({\bar x}_2 P_2) \gamma^\alpha (\not\! P_2-x_1\not\!
P_1+\not l)\gamma_\mu (\not n_+\!\!\not n_-\!\!-1)\gamma_5]}
{(P_2-x_1P_1+l)^2(x_2P_2-x_1P_1+l)^2}\;. \label{2bc}
\end{eqnarray}
To derive the above expression, the twist-2 structure
$(\gamma_5\not\! n_-)_{ik}(\not\! n_+\gamma_5)_{jl}$ has been dropped.
The dependence on $l^-$ and on $l_T$ in $H_S^{(0)}$ and in
$H_T^{(0)}$, being subleading according to Eq.~(\ref{sog2}), needs
to be neglected. Inserting the identity $\int
d\xi_1\delta(\xi_1-x_1+l^+/P_1^+)$, the first factors of the above
two terms on the right-hand side of Eq.~(\ref{2bc}) give the
collinear divergent pieces $\phi^{(1)}_{Sb}(x_1,\xi_1)$ and
$\phi^{(1)}_{Tb}(x_1,\xi_1)$ defined in Eq.~(\ref{p2b}). The
collinear factorization of Figs.~2(a) and 2(c) can be performed in
a similar way, leading to Eqs.~(\ref{p2a}) and (\ref{p2c}).

The loop integral from Fig.~2(d) is written as
\begin{eqnarray}
I^{(d)}&=& \frac{ieg^4}{2N_c} \int\frac{d^4l}{(2\pi)^4}
{\bar d}(x_1P_1)\gamma^\lambda
{\bar d}(x_2 P_2)u({\bar x}_2 P_2)
\gamma^\beta\frac{\not\! P_2-x_1\not\! P_1+\not l}{(P_2-x_1P_1+l)^2}
\gamma_\mu\frac{{\bar x}_1\not\! P_1 +\not l}{({\bar x}_1 P_1 +l)^2}
\nonumber\\
& &\times \gamma^\alpha u({\bar x}_1P_1)
\frac{{\rm tr}(T^cT^bT^a)\Gamma^{cba}_{\lambda\beta\alpha}}
{l^2(x_1P_1-x_2P_2-l)^2(x_1P_1-x_2P_2)^2}\;,
\label{2d0}
\end{eqnarray}
with the color matrices $T^{a,b,c}$ and the triple-gluon vertex,
\begin{eqnarray}
\Gamma^{cba}_{\lambda\beta\alpha}&=&f^{cba}
[g_{\alpha\beta}(2l-x_1P_1+x_2P_2)_\lambda
+g_{\beta\lambda}(2x_1P_1-2x_2P_2-l)_\alpha
\nonumber\\
& &+g_{\lambda\alpha}(x_2P_2-x_1P_1-l)_\beta]\;,
\label{tg}
\end{eqnarray}
$f^{abc}$ being an antisymmetric tensor. The above color structure
can be simplified by employing the identities,
\begin{eqnarray}
tr(T^aT^bT^c)=\frac{1}{4}(d^{abc}+if^{abc})\;,\;\;\;
d^{abc}f^{abc}=0\;,\;\;\;f^{abc}f^{abc}=24\;, \label{tco}
\end{eqnarray}
$d^{abc}$ being a symmetric tensor.

In the collinear region with $l$ parallel to $P_1$, the terms
proportional to $g_{\alpha\beta}$ and $g_{\lambda\alpha}$ do not
contribute. Since the gamma matrices must be
$\gamma^\alpha=\gamma^+$, $\gamma^\beta=\gamma^-$, and
$\gamma_\mu=\gamma_-=\gamma^+$, the quark propagator between
$\gamma_\mu$ and $\gamma^\beta$ is proportional to $l_T$, which is
subleading in the collinear region. The factor $g_{\lambda\alpha}$
indicates that $\gamma^\lambda$ must be $\gamma^-$, because of
$\gamma^\alpha=\gamma^+$. According to Eq.~(\ref{eqm}), the
contribution from ${\bar d}(x_1P_1)\gamma^\lambda={\bar
d}(x_1P_1)\gamma^-$ vanishes. The second term associated with
$g_{\beta\lambda}$ in Eq.~(\ref{tg}) contains a collinear
divergence. Due to $\gamma^\alpha=\gamma^+$, only the term
$-2x_2P_{2\alpha}$ contributes. In the collinear region we have
the approximation,
\begin{eqnarray}
\frac{-2x_2P_{2\alpha}}{(x_1P_1-x_2P_2)^2(x_1P_1-x_2P_2-l)^2}
\approx -\frac{n_{-\alpha}}{n_-\cdot l}\biggl[\frac{1}{(x_1P_1-x_2P_2)^2}
-\frac{1}{(x_1P_1-x_2P_2-l)^2}\biggr]\;.
\label{pi}
\end{eqnarray}
The expression in Eq.~(\ref{2d0}) is then split, after the
insertion of Eq.~(\ref{1f}), into two terms as shown in
Eq.~(\ref{p2d}).

The loop integral associated with Fig.~2(e) is given by
\begin{eqnarray}
I^{(e)}&=&\frac{-eg^4 C_F}{4N_c}\int\frac{d^4l}{(2\pi)^4}
{\bar d}(x_1P_1)\gamma_\alpha
{\bar d}(x_2 P_2)u({\bar x}_2 P_2)
\gamma_\beta\frac{{\bar x}_2\not\! P_2+\not l}{({\bar x}_2P_2+l)^2}
\gamma^\alpha\frac{\not\! P_2-x_1\not\! P_1 +\not l}{(P_2-x_1 P_1 +l)^2}
\gamma_\mu
\nonumber\\
& &\times \frac{{\bar x}_1\not\! P_1 +\not l}{({\bar x}_1P_1+l)^2}
\gamma^\beta u({\bar x}_1P_1)\frac{1}{l^2(x_1P_1-x_2P_2)^2}\;.
\label{2e}
\end{eqnarray}
In the collinear region $\gamma^\beta$, $\gamma_\beta$,
$\gamma^\alpha$ and $\gamma_\alpha$ must be $\gamma^+$, $\gamma^-$,
$\gamma^T$ and $\gamma^T$, respectively. Using the eikonal approximation,
\begin{eqnarray}
u({\bar x}_2P_2)\gamma_\beta
\frac{{\bar x}_2\not\! P_2+\not l}{({\bar x}_2P_2+l)^2}
\approx u({\bar x}_2P_2)\frac{n_{-\beta}}{n_-\cdot l}\;,
\label{eap}
\end{eqnarray}
and inserting Eq.~(\ref{1f}), Eq.~(\ref{2e}) leads to Eq.~(\ref{p2e}).

Following the similar treatment, the loop integral associated with
Fig.~2(f) reduces to
\begin{eqnarray}
I^{(f)}&=&\frac{eg^4 C_F}{4N_c}\int\frac{d^4l}{(2\pi)^4}
{\bar d}(x_1P_1)\gamma^\alpha
\frac{x_2\not\! P_2+\not l}{(x_2P_2+l)^2}\gamma_\beta
{\bar d}(x_2 P_2)u({\bar x}_2 P_2)
\gamma_\alpha\frac{\not\! P_2-x_1\not\! P_1 +\not l}{(P_2-x_1 P_1 +l)^2}
\gamma_\mu
\nonumber\\
& &\times \frac{\bar{x}_1\not\! P_1 +\not l}{(\bar{x}_1P_1+l)^2}
\gamma^\beta u({\bar x}_1P_1)\frac{1}{l^2(x_1P_1-x_2P_2-l)^2}\;,
\end{eqnarray}
which then gives Eq.~(\ref{p2f}).

The integral of Fig.~2(g) is written as,
\begin{eqnarray}
I^{(g)}&=&\frac{eg^4 C_F^2}{2} \int\frac{d^4l}{(2\pi)^4}
{\bar d}(x_1P_1)\gamma_\alpha
{\bar d}(x_2 P_2)u({\bar x}_2 P_2)
\gamma^\alpha\frac{\not\! P_2-x_1\not\! P_1}{(P_2-x_1P_1)^2}\gamma_\beta
\frac{\not\! P_2-x_1\not\! P_1+\not l}{(P_2-x_1P_1+l)^2}\gamma_\mu
\nonumber\\
& &\times \frac{\bar{x}_1\not\! P_1 +\not l}{(\bar{x}_1P_1 +l)^2}
\gamma^\beta u({\bar x}_1P_1)\frac{1}{l^2(x_1P_1-x_2P_2)^2}\;.
\label{2g}
\end{eqnarray}
In the collinear region with $l$ parallel to $P_1$, we have the
sequence of the gamma matrices $\gamma^\beta=\gamma^+$,
$\gamma_\mu=\gamma^+$, and $\gamma_\beta=\gamma^-$. The quark
propagator between $\gamma_\mu$ and $\gamma_\beta$ gives the
subleading contribution proportional to $l_T$. Hence, Fig.~2(g)
does not contribute a collinear divergence.

Figure 2(h) gives the loop integral,
\begin{eqnarray}
I^{(h)}&=&\frac{-ieg^4}{2N_c}\int\frac{d^4l}{(2\pi)^4}
{\bar d}(x_1P_1)\gamma^\lambda
\frac{x_1\not\! P_1 -\not l}{(x_1 P_1 -l)^2}
\gamma^\beta{\bar d}(x_2P_2)u({\bar x}_2P_2)\gamma^\alpha
\frac{\not\! P_2-x_1\not\! P_1}{(P_2-x_1P_1)^2}
\nonumber\\
& &\times \gamma_\mu u({\bar x}_1P_1)
\frac{{\rm tr}(T^cT^bT^a)\Gamma^{cba}_{\lambda\beta\alpha}}
{l^2(x_1P_1-x_2P_2-l)^2(x_1P_1-x_2P_2)^2}\;,
\label{2h0}
\end{eqnarray}
with the triple-gluon vertex,
\begin{eqnarray}
\Gamma^{cba}_{\lambda\beta\alpha}&=&f^{cba}
[g_{\beta\lambda}(2l-x_1P_1+x_2P_2)_\alpha
+g_{\alpha\beta}(2x_1P_1-2x_2P_2-l)_\lambda
\nonumber\\
& &+g_{\lambda\alpha}(x_2P_2-x_1P_1-l)_\beta]\;.
\label{tge}
\end{eqnarray}
In the collinear region with $l$ parallel to $P_1$, the terms
proportional to $g_{\beta\lambda}$ and $g_{\lambda\alpha}$ do not
contribute. Because $\gamma^\lambda$ must be $\gamma^+$, the
factor $g_{\beta\lambda}$ implies that $\gamma^\beta$ must be
$\gamma^-$ and the quark propagator between $\gamma^\lambda$ and
$\gamma^\beta$ is proportional to $l_T$, which is subleading. The
factor $g_{\lambda\alpha}$ implies $\gamma^\alpha =\gamma^-$.
Since $\gamma_\mu$ is adjacent to the spinor $u(\bar{x}_1P_1)$, it
must be $\gamma^+$. The quark propagator between $\gamma_\mu$ and
$\gamma^\alpha$ gives the subleading contribution proportional to
$l_T$. Only the second term in Eq.~(\ref{tge}) contains a
collinear divergence.

Applying the approximation in Eq.~(\ref{pi}), and inserting
Eq.~(\ref{1f}), Eq.~(\ref{2h0}) becomes
\begin{eqnarray}
I^{(h)}\approx \sum_{n=S,T} \int d \xi_1 \phi_{nh}^{(1)}(x_1, \xi_1)
H_n^{(0)}(\xi_1, x_2)\;,
\label{2h2}
\end{eqnarray}
with the $O(\alpha_s)$ collinear divergent PS piece,
\begin{eqnarray}
\phi_{Sh}^{(1)}(x_1, \xi_1)&=&\frac{ig^2}{2m_0C_F}
\int\frac{d^4l}{(2\pi)^4}{\bar d}(x_1P_1)\gamma^\beta
\frac{x_1\not\! P_1 -\not l}{(x_1 P_1 -l)^2}
\gamma^5u({\bar x}_1P_1)\frac{1}{l^2}
\frac{n_{-\beta}}{n_-\cdot l}
\nonumber \\
& &\times\left[\delta(\xi_1-x_1)-
\delta\left(\xi_1-x_1+\frac{l^+}{P_1^+}\right)\right]\;.
\label{p2h}
\end{eqnarray}

The loop integral associated with Fig.~2(i) is written as
\begin{eqnarray}
I^{(i)}&=&\frac{-eg^4 C_F}{4N_c} \int\frac{d^4l}{(2\pi)^4}
{\bar d}(x_1P_1)\gamma^\alpha
\frac{x_1\not\! P_1-\not l}{(x_1P_1-l)^2}\gamma^\beta
\frac{x_2\not\! P_2-\not l}{(x_2P_2-l)^2}\gamma_\alpha{\bar d}(x_2P_2)
\nonumber \\
& &\times u({\bar x}_2P_2)\gamma_\beta
\frac{\not\! P_2-x_1\not\! P_1}{(P_2-x_1 P_1)^2}
\gamma_\mu u({\bar x}_1P_1)\frac{1}{l^2(x_1P_1-x_2P_2)^2}\;.
\label{2i}
\end{eqnarray}
After applying the eikonal approximation similar to
Eq.~(\ref{eap}), and inserting Eq.~(\ref{1f}), the above
expression is simplified into
\begin{eqnarray}
I^{(i)}\approx  \sum_{n=S,T}\int d\xi_1
\phi_{ni}^{(1)}(x_1, \xi_1)H_n^{(0)}(\xi_1,x_2)\;,
\label{2i2}
\end{eqnarray}
with the collinear divergent PS piece,
\begin{eqnarray}
\phi_{Si}^{(1)}(x_1, \xi_1)=\frac{-ig^2}{8m_0N_c}
\int\frac{d^4l}{(2\pi)^4}{\bar d}(x_1P_1)\gamma^\beta
\frac{x_1\not\! P_1 -\not l}{(x_1 P_1 -l)^2}
\gamma^5u({\bar x}_1P_1)\frac{1}{l^2}
\frac{n_{-\beta}}{n_-\cdot l}\delta (\xi_1 - x_1)\;.
\label{p2i}
\end{eqnarray}

Figure 2(j) gives the loop integral,
\begin{eqnarray}
I^{(j)}&=&\frac{eg^4 C_F}{4N_c}\int\frac{d^4l}{(2\pi)^4}
{\bar d}(x_1P_1)\gamma^\alpha
\frac{x_1\not\! P_1-\not l}{(x_1P_1-l)^2}\gamma^\beta{\bar d}(x_2P_2)
u({\bar x}_2P_2)\gamma_\alpha
\frac{\bar{x}_2\not\! P_2-\not l}{(\bar{x}_2P_2-l)^2}
\nonumber \\
& &\times \gamma_\beta\frac{\not\! P_2-x_1\not\! P_1}{(P_2-x_1 P_1)^2}
\gamma_\mu u({\bar x}_1P_1)\frac{1}{l^2(x_1P_1-x_2P_2-l)^2}\;.
\label{2j}
\end{eqnarray}
Following the same procedure as for Eq.~(\ref{2i}), we obtain
\begin{eqnarray}
I^{(j)}\approx \sum_{n=S,T}\int d\xi_1 \phi_{nj}^{(1)}(x_1, \xi_1)H_n^{(0)}(\xi_1,x_2)\;,
\label{2j2}
\end{eqnarray}
with the collinear divergent PS piece, 
\begin{eqnarray}
\phi_{Sj}^{(1)}(x_1, \xi_1)=\frac{ig^2}{8m_0N_c}
\int\frac{d^4l}{(2\pi)^4}{\bar d}(x_1P_1)\gamma^\beta
\frac{x_1\not\! P_1 -\not l}{(x_1 P_1 -l)^2}
\gamma^5u({\bar x}_1P_1)\frac{1}{l^2}
\frac{n_{-\beta}}{n_-\cdot l}\delta\left(\xi_1 - x_1 + \frac{l^+}{P_1^+}\right)\;.
\label{p2j}
\end{eqnarray}

At last, the integral associated with Fig.~2(k) is given by
\begin{eqnarray}
I^{(k)}&=&\frac{-eg^4 C_F^2}{2}\int\frac{d^4l}{(2\pi)^4}
{\bar d}(x_1P_1)\gamma^\alpha
\frac{x_1\not\! P_1 -\not l}{(x_1 P_1 -l)^2}
\gamma^\beta{\bar d}(x_2P_2)u({\bar x}_2P_2)\gamma_\beta
\frac{\not\! P_2-x_1\not\! P_1+\not l}{(P_2-x_1P_1+l)^2}\gamma_\alpha
\nonumber \\
& &\times \frac{\not\! P_2-x_1\not\! P_1}{(P_2-x_1P_1)^2}
\gamma_\mu u({\bar x}_1P_1)\frac{1}{l^2(x_1P_1-x_2P_2-l)^2}\;.
\label{2k}
\end{eqnarray}
It is easy to find that a collinear divergence does not exist for
the same reason as for Fig.~2(g): in the collinear region we have
the gamma matrices $\gamma^\alpha=\gamma^+$ and
$\gamma^\beta=\gamma^+$, which is adjacent to the the spinor
${\bar d}(x_2P_2)$. The contribution is then subleading because of
the equations of motion for the final-state quarks.

\section{$O(\alpha_s)$ CORRECTIONS FROM FIG.~4}

In this Appendix we discuss the factorization of the initial-state
collinear divergences in the $O(\alpha_s)$ radiative corrections
to Fig.~1(b), which are shown in Fig.~4. Figure 1(b) gives only
the lowest-order PS distribution amplitude $\phi_S^{(0)}$ and the
hard amplitude $H_S^{(0)}$,
\begin{eqnarray}
& &\phi_S^{(0)}(x_1,\xi_1)=\frac{1}{4m_0}{\bar d}(x_1P_1)\gamma_5
u({\bar x}_1P_1)\delta(\xi_1-x_1)\;,
\nonumber\\
& &H_{S}^{(0)}(\xi_1,x_2)=\frac{i}{2}eg^2 C_Fm_0
\frac{{\rm tr}[\gamma_\nu
{\bar d}(x_2P_2)u({\bar x}_2P_2)\gamma_\mu
(\not\! P_1-x_2\not\! P_2)\gamma^\nu\gamma^5]}
{(P_1-x_2P_2)^2(\xi_1P_1-x_2P_2)^2}\;,
\label{4bc}
\end{eqnarray}
because of $\gamma^\nu (\not n_+\not n_--1)\gamma_\nu=0$,
where the gamma matrices $\gamma^\nu$ and $\gamma_\nu$
come from the gluon vertices in Fig.~1(b).

The integral from Fig.~4(b) is written as
\begin{eqnarray}
I^{(b)}&=& -\frac{1}{2}eg^4 C_F^2\int\frac{d^4l}{(2\pi)^4} {\bar
d}(x_1P_1)\gamma_\beta \frac{x_1\not\! P_1-\not
l}{(x_1P_1-l)^2}\gamma_\alpha {\bar d}(x_2P_2)u({\bar x}_2P_2)
\nonumber\\
& &\times
\gamma_\mu\frac{\not\! P_1-x_2\not\! P_2}{(P_1-x_2P_2)^2}
\gamma^\alpha \frac{{\bar x}_1\not\! P_1 +\not l}{({\bar x}_1 P_1 +l)^2}
\gamma^\beta u({\bar x}_1P_1)\frac{1}{l^2(x_2P_2-x_1P_1+l)^2}\;.
\label{4b}
\end{eqnarray}
Following the same procedure as for Eq.~(\ref{2b}), we obtain
\begin{eqnarray}
I^{(b)}&\approx& \int d\xi_1 \phi^{(1)}_{Sb}(x_1,\xi_1)
H_S^{(0)}(\xi_1,x_2)\;, \label{4b1}
\end{eqnarray}
where the expression of $\phi_{Sb}^{(1)}(x_1,\xi_1)$ has been
given in Eq.~(\ref{p2b}). The factorization of Figs.~4(a) and 4(c)
is performed in a similar way, leading to
$\phi_{Sa,Sc}^{(1)}(x_1,\xi_1)$ in Eqs.~(\ref{p2a}) and
(\ref{p2c}).

The loop integral associated with Fig.~4(d) is given by
\begin{eqnarray}
I^{(d)}&=&\frac{ieg^4}{2N_c}\int\frac{d^4l}{(2\pi)^4} {\bar
d}(x_1P_1)\gamma^\lambda{\bar d}(x_2P_2)u({\bar x}_2P_2)
\gamma_\mu\frac{\not\! P_1-x_2\not\! P_2}{(P_1-x_2P_2)^2}\gamma^\beta
\frac{{\bar x}_1\not\! P_1 +\not l}{({\bar x}_1 P_1 +l)^2}
\nonumber\\
& &\times \gamma^\alpha u({\bar x}_1P_1)
\frac{{\rm tr}(T^cT^bT^a)\Gamma^{cba}_{\lambda\beta\alpha}}
{l^2(x_1P_1-x_2P_2-l)^2(x_1P_1-x_2P_2)^2}\;,
\label{4d}
\end{eqnarray}
with the triple-gluon vertex in Eq.~(\ref{tg}). The same procedure as for
Eq.~(\ref{2d0}) leads to
\begin{eqnarray}
I^{(d)} &\approx & \int d\xi_1 \phi_{Sd}^{(1)}(x_1,\xi_1)
H_S^{(0)}(\xi_1, x_2)\;,
\label{4d1}
\end{eqnarray}
with the function $\phi_{Sd}^{(1)}(x_1,\xi_1)$ shown in
Eq.~(\ref{p2d}).

The loop integral associated with Fig.~4(e) is written as
\begin{eqnarray}
I^{(e)}&=&\frac{-eg^4 C_F}{4N_c}\int\frac{d^4l}{(2\pi)^4} {\bar
d}(x_1P_1)\gamma^\alpha{\bar d}(x_2P_2)u({\bar
x}_2P_2)\gamma_\beta \frac{\bar{x}_2\not\! P_2+\not
l}{(\bar{x}_2P_2+l)^2} \gamma_\mu\frac{\not\! P_1-x_2\not\! P_2+\not
l}{(P_1-x_2 P_2+l)^2} \gamma_\alpha
\nonumber \\
& &\times
\frac{\bar{x}_1\not\! P_1 +\not l}{(\bar{x}_1P_1+l)^2}
\gamma^\beta u({\bar x}_1P_1)\frac{1}{l^2(x_1P_1-x_2P_2)^2}\;,
\label{4e}
\end{eqnarray}
which is simplified into
\begin{eqnarray}
I^{(e)}&\approx& \frac{ig^2}{8m_0N_c}
\int\frac{d^4l}{(2\pi)^4}{\bar d}(x_1P_1)\gamma_5 \frac{{\bar
x}_1\not\! P_1 +\not l}{({\bar x}_1 P_1 +l)^2}\gamma^\beta u({\bar
x}_1P_1)\frac{1}{l^2} \frac{n_{-\beta}}{n_-\cdot l} \frac{(P_1-x_2
P_2)^2}{(P_1-x_2 P_2+l)^2} H_{S}^{(0)}(\xi_1,x_2) \;.
\label{4e1}
\end{eqnarray}

On the other hand,
the loop integral associated with Fig.~4(g) is written as
\begin{eqnarray}
I^{(g)}&=&\frac{-eg^4 C_F}{4N_c}\int\frac{d^4l}{(2\pi)^4} {\bar
d}(x_1P_1)\gamma^\alpha{\bar d}(x_2P_2)u({\bar x}_2P_2)\gamma_\mu
\frac{\not\! P_1-x_2\not\! P_2}{(P_1-x_2P_2)^2}\gamma_\beta \frac{\not\!
P_1-x_2\not\! P_2+\not l}{(P_1-x_2P_2+l)^2} \gamma_\alpha
\nonumber \\
& &\times
\frac{\bar{x}_1\not\! P_1+l}{(\bar{x}_1P_1+l)^2}
\gamma^\beta u({\bar x}_1P_1)\frac{1}{l^2(x_1P_1-x_2P_2)^2}\;,
\label{4g}
\end{eqnarray}
which reduces to
\begin{eqnarray}
I^{(g)}&\approx& \frac{ig^2}{8m_0N_c}
\int\frac{d^4l}{(2\pi)^4}{\bar d}(x_1P_1)\gamma_5 \frac{{\bar
x}_1\not\! P_1 +\not l}{({\bar x}_1 P_1 +l)^2}\gamma^\beta u({\bar
x}_1P_1)\frac{1}{l^2} \frac{n_{-\beta}}{n_-\cdot
l}\frac{-2x_2P_2\cdot l}{(P_1-x_2 P_2+l)^2}
H_{S}^{(0)}(\xi_1,x_2)\;. \label{4g1}
\end{eqnarray}
This result differs from that of Fig.~2(g), which does not contain
a collinear divergence. Neither Eq.~(\ref{4e1}) nor (\ref{4g1}) is
in the desired form. However, their combination is, as shown
below,
\begin{eqnarray}
I^{(e)}+I^{(g)}&\approx&
\int d\xi_1 \phi_{Se}^{(1)}(x_1,\xi_1)H_{S}^{(0)}(\xi_1,x_2)\;,
\label{4eg}
\end{eqnarray}
with the function $\phi^{(1)}_{Se}(x_1,\xi_1)$ shown in  Eq.~(\ref{p2e}).

The loop integral associated with Fig.~4(f),
\begin{eqnarray}
I^{(f)}&=&\frac{eg^4 C_F}{4N_c}\int\frac{d^4l}{(2\pi)^4} {\bar
d}(x_1P_1)\gamma^\alpha \frac{x_2\not\! P_2+\not l}{(x_2P_2+l)^2}
\gamma_\beta{\bar d}(x_2P_2)u({\bar x}_2P_2) \gamma_\mu\frac{\not\!
P_1-x_2\not\! P_2}{(P_1-x_2 P_2)^2} \gamma_\alpha
\nonumber \\
& &\times
\frac{\bar{x}_1\not\! P_1 +\not l}{(\bar{x}_1P_1+l)^2}
\gamma^\beta u({\bar x}_1P_1)\frac{1}{l^2(x_1P_1-x_2P_2-l)^2}\;,
\label{4f}
\end{eqnarray}
leads to
\begin{eqnarray}
I^{(f)}&\approx& \int d\xi_1 \phi_{Sf}^{(1)}(x_1,\xi_1)
H_{S}^{(0)}(\xi_1,x_2)\;,
\label{4f1}
\end{eqnarray}
where the function $\phi^{(1)}_{Sf}(x_1,\xi_1)$ has been given in
Eq.~(\ref{p2f}).

The integral associated with Fig.~4(h) is given by
\begin{eqnarray}
I^{(h)}&=&\frac{-ieg^4}{2N_c}\int\frac{d^4l}{(2\pi)^4} {\bar
d}(x_1P_1)\gamma^\lambda \frac{x_1\not\! P_1 -\not l}{(x_1 P_1
-l)^2} \gamma^\beta{\bar d}(x_2P_2)u({\bar x}_2P_2)\gamma_\mu
\frac{\not\! P_1-x_2\not\! P_2}{(P_1-x_2P_2)^2}
\nonumber\\
& &\times \gamma^\alpha u({\bar x}_1P_1)
\frac{{\rm tr}(T^cT^bT^a)\Gamma^{cba}_{\lambda\beta\alpha}}
{l^2(x_1P_1-x_2P_2-l)^2(x_1P_1-x_2P_2)^2}\;,
\label{4h}
\end{eqnarray}
with the triple-gluon vertex in Eq.~(\ref{tge}). Following the
same procedure as for Fig.~4(d), the integral becomes
\begin{eqnarray}
I^{(h)}&\approx& \int d\xi_1 \phi_{Sh}^{(1)}(x_1,\xi_1)
H_{S}^{(0)}(\xi_1,x_2)\;.
\label{4h1}
\end{eqnarray}
Similarly, we obtain
\begin{eqnarray}
I^{(i)}&\approx&
\int d\xi_1 \phi_{Si}^{(1)}(x_1,\xi_1)H_{S}^{(0)}(\xi_1,x_2)\;,
\label{4i1}
\\
I^{(j)}+I^{(k)}&\approx&
\int d\xi_1 \phi_{Sj}^{(1)}(x_1,\xi_1)H_{S}^{(0)}(\xi_1,x_2)\;.
\label{4jk}
\end{eqnarray}
The functions $\phi^{(1)}_{Sh}(x_1,\xi_1)$,
$\phi^{(1)}_{Si}(x_1,\xi_1)$ and
$\phi^{(1)}_{Sj}(x_1,\xi_1)$
have been presented in Eqs.~(\ref{p2h}), (\ref{p2i}) and (\ref{p2j})
respectively.

Combining all the above contributions from Figs.~4(a)-4(k), we
derive the $O(\alpha_s)$ factorization in Eq.~(\ref{phi3}).

\section{$O(\alpha_s)$ CORRECTIONS TO $B\to \pi \ell\bar \nu$}

In this Appendix we discuss the factorization of the collinear
divergences in Fig.4 for the semileptonic decay $B\to\pi l\bar\nu$
with the initial state and the final state being flipped. Figures
4(a)-4(c) can be factorized straightforwardly, leading to
Eq.~(\ref{4a2}). Note that the hard amplitude from Fig.~4(b)
depends on the loop momentum $l$.

The loop integral from Fig.~4(d) is given by
\begin{eqnarray}
I^{(d)}&=&\frac{-g^4}{2N_c}\int\frac{d^4l}{(2\pi)^4} u({\bar
x}_2P_2)\gamma^\beta \frac{{\bar x}_2\not\! P_2+\not l}{({\bar
x}_2P_2+l)^2}\gamma^\alpha \frac{\not\! P_2-\not
k_1}{(P_2-k_1)^2}\gamma_\mu b(P_1-k_1){\bar d}(k_1)
\nonumber\\
& &\times\gamma^\lambda {\bar d}(x_2P_2)
\frac{tr(T^cT^bT^a)\Gamma^{cba}_{\lambda\beta\alpha}}
{l^2(k_1-x_2P_2+l)^2(k_1-x_2P_2)^2}\;,
\label{4bd0}
\end{eqnarray}
with the triple-gluon vertex,
\begin{eqnarray}
\Gamma^{cba}_{\lambda\beta\alpha}&=&f^{cba}
[g_{\alpha\beta}(2l+k_1-x_2P_2)_\lambda
+g_{\beta\lambda}(k_1-x_2P_2-l)_\alpha
\nonumber\\
& &+g_{\lambda\alpha}(2x_2P_2-2k_1-l)_\beta]\;.
\end{eqnarray}
The color factor is simplified according to Eq.~(\ref{tco}).

In the collinear region with $l$ parallel to $P_2$, only the term
proportional to $g_{\alpha\lambda}$ contributes by employing the
argument the same as for Fig.~2(d) in Appendix A. Since
$\gamma^\beta$ must be $\gamma^-$, only the plus component of
$k_1$ survives. Applying the approximation similar to
Eq.~(\ref{pi}),
\begin{eqnarray}
\frac{-2k_{1\beta}}{(k_1-x_2P_2)^2(k_1-x_2P_2+l)^2}
\approx -\frac{n_{+\beta}}{n_+\cdot l}\biggl[\frac{1}{(k_1-x_2P_2)^2}
-\frac{1}{(k_1-x_2P_2+l)^2}\biggr]\;,
\end{eqnarray}
and inserting Eq.~(\ref{1f}), we obtain Eq.~(\ref{pb4d}).

The collinear factorization of Fig.~4(e) can be achieved by applying the
eikonal approximation to the $b$ quark propagator,
\begin{eqnarray}
\frac{\not\! P_1-\not k_1+\not l+m_b}{(P_1-k_1+l)^2-m_b^2}
\gamma_\beta b(P_1-k_1)
&\approx&\frac{2(P_1-k_1)_\beta
-\gamma_\beta(\not\! P_1-\not k_1-m_b)}
{2(P_1-k_1)\cdot l}b(P_1-k_1)
\nonumber \\
&\approx &\frac{n_{+\beta}}{n_+\cdot l}b(P_1-k_1)\;.
\end{eqnarray}
The neglect of $\not l$ is due to $\gamma_\beta=\gamma^+$ in the
collinear region. The second term on the right-hand side of the
first line vanishes because of Eq.~(\ref{eqb}). To derive the
final expression, we have further dropped the power-suppressed
terms proportional to $k_1$. The integral associated with
Fig.~4(e) then becomes
\begin{eqnarray}
I^{(e)}&\approx& \frac{ig^2}{8m_0N_c} \int\frac{d^4l}{(2\pi)^4}
u({\bar x}_2P_2)\gamma^\beta \frac{{\bar x}_2\not\! P_2+\not
l}{({\bar x}_2P_2+l)^2}\gamma_5 {\bar
d}(x_2P_2)\frac{1}{l^2}\frac{n_{+\beta}}{n_+\cdot l}
\nonumber \\
& &\times \frac{(P_2-k_1)^2}{(P_2-k_1+l)^2}
\left(\frac{-g^2}{2}C_Fm_0\right)\frac{tr[\gamma^\alpha(\not\! P_2-\not k_1)
\gamma_\mu b(P_1-k_1){\bar d}(k_1)\gamma_\alpha\gamma^5]}
{(P_2-k_1)^2(k_1-x_2P_2)^2}\;.
\label{4be}
\end{eqnarray}
Note that the factor $(P_2-k_1)^2/(P_2-k_1+l)^2$ in the second
line indicates that Eq.~(\ref{4be}) has not yet reached the
expected factorization form.

The integral for Fig.~4(g) reduces, in a similar way, to
\begin{eqnarray}
I^{(g)}&\approx& \frac{ig^2}{8m_0N_c}\int\frac{d^4l}{(2\pi)^4}
u({\bar x}_2P_2)\gamma^\beta \frac{{\bar x}_2\not\! P_2+\not
l}{({\bar x}_2P_2+l)^2}\gamma_5 {\bar
d}(x_2P_2)\frac{1}{l^2}\frac{n_{+\beta}}{n_+\cdot l}
\nonumber \\
& &\times \frac{-2k_1\cdot l}{(P_2-k_1+l)^2}
\left(\frac{-g^2}{2}C_Fm_0\right)\frac{tr[\gamma^\alpha(\not\! P_2-\not\! k_1)
\gamma_\mu b(P_1-k_1){\bar d}(k_1)\gamma_\alpha\gamma^5]}
{(P_2-k_1)^2(k_1-x_2P_2)^2}\;.
\label{4bg}
\end{eqnarray}
Combining Eqs.~(\ref{4be}) and (\ref{4bg}), we arrive at the
desired factorization form in Eq.~(\ref{4f2}) with the collinear
piece shown in Eq.~(\ref{pb4eg}). It is observed that the hard
part is independent of the loop momentum $l$.

The integral associated with Fig.~4(f) is factorized into
Eq.~(\ref{4f2}) straightforwardly with the collinear piece shown
in Eq.~(\ref{pb4f}). In this case the eikonal propagator arises
from the approximation,
\begin{eqnarray}
{\bar d}(k_1)\gamma_\beta
\frac{\not k_1+\not l}{(k_1+l)^2}
\approx {\bar d}(k_1)\frac{2k_{1\beta}-\not\! k_1\gamma_\beta}
{2k_1\cdot l}
={\bar d}(k_1)\frac{n_{+\beta}}{n_+\cdot l}\;.
\label{bfap}
\end{eqnarray}
Again, the neglect of $\not l$ is due to $\gamma_\beta=\gamma^+$ in the
collinear region, and Eq.~(\ref{eqb}) has been employed to derive the
final expression.

The loop integral from Fig.~4(h) is written as
\begin{eqnarray}
I^{(h)}&=&\frac{g^4}{2N_c}\int\frac{d^4l}{(2\pi)^4} u({\bar
x}_2P_2)\gamma^\alpha \frac{\not\! P_2-\not\!
k_1}{(P_2-k_1)^2}\gamma_\mu b(P_1-k_1){\bar d}(k_1) \gamma^\lambda
\frac{x_2\not\! P_2-\not l}{(x_2P_2-l)^2}
\nonumber\\
& &\times\gamma^\beta{\bar d}(x_2P_2)
\frac{tr(T^cT^bT^a)\Gamma^{cba}_{\lambda\beta\alpha}}
{l^2(k_1-x_2P_2+l)^2(k_1-x_2P_2)^2}\;,
\label{4bh0}
\end{eqnarray}
with the triple-gluon vertex,
\begin{eqnarray}
\Gamma^{cba}_{\lambda\beta\alpha}&=&f^{cba}
[g_{\beta\lambda}(2l+k_1-x_2P_2)_\alpha
+g_{\alpha\beta}(k_1-x_2P_2-l)_\lambda
\nonumber\\
& &+g_{\lambda\alpha}(2x_2P_2-2k_1-l)_\beta]\;.
\end{eqnarray}
Following the same procedure as for Fig.~4(d), Eq.~(\ref{4bh0})
reduces to
\begin{eqnarray}
I^{(h)}\approx \int d\xi_2
H_S^{(1)}(x_1,\xi_2)\phi_{Sh}^{(1)}(x_2,\xi_2)\;,
\label{4bh2}
\end{eqnarray}
with the $O(\alpha_s)$ collinear divergent function,
\begin{eqnarray}
\phi_{Sh}^{(1)}(x_2,\xi_2)&=&\frac{ig^2}{2m_0C_F}\int\frac{d^4l}{(2\pi)^4}
u({\bar x}_2P_2)\gamma_5
\frac{x_2\not\! P_2 -\not l}{(x_2 P_2 -l)^2}
\gamma^\beta{\bar d}(x_2P_2)\frac{1}{l^2}
\frac{n_{+\beta}}{n_+\cdot l}
\nonumber \\
& &\times \left[\delta(\xi_2-x_2)-
\delta\left(\xi_2-x_2+\frac{l^-}{P_2^-}\right)\right] \;.
\label{p4bh}
\end{eqnarray}

The loop integral associated with Fig.~4(i) is given by
\begin{eqnarray}
I^{(i)}&=&\frac{-ig^4 C_F}{4N_c}\int\frac{d^4l}{(2\pi)^4} u({\bar
x}_2P_2)\gamma^\alpha \frac{\not\! P_2-\not\!
k_1}{(P_2-k_1)^2}\gamma_\mu b(P_1-k_1){\bar d}(k_1)\gamma_\beta
\frac{\not\! k_1-\not l}{(k_1-l)^2}\gamma_\alpha
\nonumber \\
& & \times \frac{x_2\not\! P_2-\not l}{(x_2P_2-\l)^2}
\gamma^\beta{\bar d}(x_2P_2)
\frac{1}{l^2(k_1-x_2P_2)^2}\;.
\label{4bi}
\end{eqnarray}
Using the approximation similar to Eq.~(\ref{bfap}),
the above expression is factorized as
\begin{eqnarray}
I^{(i)}\approx  \int d\xi_2 H_S^{(0)}(x_1,\xi_2)\phi_{Si}^{(1)}(x_2,\xi_2)\;,
\label{4bi2}
\end{eqnarray}
with the collinear divergent function,
\begin{eqnarray}
\phi_{Si}^{(1)}(x_2,\xi_2)=\frac{-ig^2}{8m_0N_c}\int\frac{d^4l}{(2\pi)^4}
u({\bar x}_2P_2)\gamma_5
\frac{x_2\not\! P_2 -\not l}{(x_2 P_2 -l)^2}
\gamma^\beta{\bar d}(x_2P_2)\frac{1}{l^2}
\frac{n_{+\beta}}{n_+\cdot l}\delta (\xi_2-x_2)\;.
\label{p4bi}
\end{eqnarray}

Following the same procedure as for Eqs.~(\ref{4be}) and (\ref{4bg}),
Figs.~4(j) and 4(k) lead to
\begin{eqnarray}
I^{(j)}&\approx& \frac{ig^2}{8m_0N_c}\int\frac{d^4l}{(2\pi)^4}
H_{S}^{(0)}(x_1,\xi_2)u({\bar x}_2P_2)\gamma_5 \frac{x_2\not\!
P_2-\not l}{(x_2P_2-l)^2}\gamma^\nu {\bar d}(x_2P_2)\frac{1}{l^2}
\frac{n_{+\nu}}{n_+\cdot l}\frac{(P_2-k_1)^2}{(P_2-k_1-l)^2}\;,
\nonumber\\
I^{(k)}&\approx& \frac{ig^2}{8m_0N_c}\int\frac{d^4l}{(2\pi)^4}
H_{S}^{(0)}(x_1,\xi_2)u({\bar x}_2P_2)\gamma_5 \frac{x_2\not\!
P_2-\not l}{(x_2P_2-l)^2}\gamma^\nu {\bar d}(x_2P_2)\frac{1}{l^2}
\frac{n_{+\nu}}{n_+\cdot l}\frac{2k_1\cdot l}{(P_2-k_1-l)^2}\;.
\label{4bjk0}
\end{eqnarray}
Combining the expressions in Eq.~(\ref{4bjk0}), we arrive at the desired
factorization form,
\begin{eqnarray}
I^{(j)}+I^{(k)}&\approx& \int d\xi_2
H_{S}^{(0)}(x_1,\xi_2)\phi_{Sj}^{(1)}(x_2,\xi_2)\;,
\label{4bjk}
\end{eqnarray}
with the collinear divergent piece,
\begin{eqnarray}
\phi_{Sj}^{(1)}(\xi_1,x_2)=\frac{ig^2}{8m_0N_c}\int\frac{d^4l}{(2\pi)^4}
u({\bar x}_2P_2)\gamma_5 \frac{x_2\not\! P_2-\not
l}{(x_2P_2-l)^2}\gamma^\nu {\bar d}(x_2P_2)\frac{1}{l^2}
\frac{n_{+\nu}}{n_+\cdot
l}\delta\left(\xi_2-x_2+\frac{l^-}{P_2^-}\right)\;. \label{pb4j}
\end{eqnarray}

We then discuss the factorization of the collinear divergences
from Fig.~2 with the initial and final states being flipped, which
represents the $O(\alpha_s)$ corrections to Fig.~1(b). To simplify
the discussion, we show only the PS parts below. The results are
\begin{eqnarray}
I^{(d)}\!&\approx &\!\int\frac{d^4l}{(2\pi)^4} \Biggl(\frac{-g^2
C_F}{2}m_0\Biggr)\frac{tr[\gamma_\mu (\not\! P_1-x_2\not\! P_2+\not
l+m_b)\gamma^\nu b(P_1-k_1) {\bar d}(k_1)\gamma_\nu\gamma^5]}
{[(P_1-x_2P_2+l)^2-m_b^2](k_1-x_2P_2)^2}
\nonumber\\
& &\times \frac{-ig^2}{2m_0C_F}
u({\bar x}_2P_2)\gamma^\beta
\frac{{\bar x}_2\not\! P_2 +\not l}{({\bar x}_2P_2+l)^2}
\gamma_5 {\bar d}(x_2P_2)\frac{1}{l^2}
\frac{n_{+\beta}}{n_+\cdot l}
\nonumber\\
& & \!\!\!\!\!\!\!
-\int d\xi_2 H^{(0)}_{S}(x_1,\xi_2)
\left[\frac{-ig^2}{2m_0C_F}\int\frac{d^4l}{(2\pi)^4}
u({\bar x}_2P_2)\gamma^\beta
\frac{{\bar x}_2\not\! P_2 +\not l}{({\bar x}_2P_2+l)^2}\right.
\nonumber\\
& &\left.\times
\gamma_5 {\bar d}(x_2P_2)\frac{1}{l^2}
\frac{n_{+\beta}}{n_+\cdot l}\delta\Biggl(\xi_2-x_2+\frac{l^-}{P_2^-}\Biggr)\right]\;,
\label{2bd}\\
I^{(e)}\!&\approx &\!\int\frac{d^4l}{(2\pi)^4}
\Biggl(\frac{-g^2 C_F}{2}m_0\Biggr)\frac{tr[\gamma_\mu
(\not\! P_1-x_2\not\! P_2+\not l+m_b)\gamma^\nu b(P_1-k_1)
{\bar d}(k_1)\gamma_\nu\gamma^5]}
{[(P_1-x_2P_2+l)^2-m_b^2](k_1-x_2P_2)^2}
\nonumber\\
& &\times \frac{ig^2}{8m_0N_c}
u({\bar x}_2P_2)\gamma^\beta
\frac{{\bar x}_2\not\! P_2+\not l}{({\bar x}_2P_2+l)^2}\gamma_5
{\bar d}(x_2P_2)\frac{1}{l^2}
\frac{n_{+\beta}}{n_+\cdot l}\;,
\label{2be}\\
I^{(f)}\!&\approx& \! \int d\xi_2H^{(0)}_{S}(x_1,\xi_2)
\phi_{Sf}^{(1)}(x_2,\xi_2)\;.
\label{2bf}
\end{eqnarray}

Combining the above expressions, we have
\begin{eqnarray}
I^{(d)}\!+\!I^{(e)}\!+\!I^{(f)}&\approx& \!\int\frac{d^4l}{(2\pi)^4}
\Biggl(\!\frac{-g^2 C_F}{2}m_0\Biggr)\frac{tr[\gamma_\mu
(\not\! P_1-x_2\!\not\! P_2+\not l+m_b)\gamma^\nu b(P_1-k_1)
{\bar d}(k_1)\gamma_\nu\gamma^5]}
{[(P_1-x_2P_2+l)^2-m_b^2](k_1-x_2P_2)^2}
\nonumber\\
& &\times \frac{-ig^2C_F}{4m_0}
u({\bar x}_2P_2)\gamma^\beta
\frac{{\bar x}_2\not\! P_2 +\not l}{({\bar x}_2P_2 +l)^2}\gamma_5
{\bar d}(x_2P_2) \frac{1}{l^2}
\frac{n_{+\beta}}{n_+\cdot l}
\nonumber\\
& &\!\!\!\!\!\!\!
-\int d\xi_2 H^{(0)}_{S}(x_1,\xi_2)
\left[\frac{-ig^2C_F}{4m_0} \int\frac{d^4l}{(2\pi)^4}
u({\bar x}_2P_2)\gamma^\beta
\frac{{\bar x}_2\not\! P_2 +\not l}{({\bar x}_2P_2 +l)^2}\right.
\nonumber\\
& & \left.\times \gamma_5{\bar d}(x_2P_2) \frac{1}{l^2}
\frac{n_{+\beta}}{n_+\cdot l}\delta\Biggl(\xi_2-x_2+\frac{l^-}{P_2^-}\Biggr)\right]\;.
\label{2bdf}
\end{eqnarray}

Consider the loop integral associated with Fig.~2(g) under the
approximation,
\begin{eqnarray}
\frac{\not\! P_1-x_2\not\! P_2+\not l +m_b}{(P_1-x_2P_2+l)^2-m_b^2}
\gamma_\beta\frac{\not\! P_1-x_2\not\! P_2+m_b}{(P_1-x_2P_2)^2-m_b^2}
&\approx&\frac{2P_{1\beta}}{(P_1-x_2P_2+l)^2-m_b^2}
\frac{\not\! P_1+m_b}{(P_1-x_2P_2)^2-m_b^2}
\nonumber \\
&\approx &\frac{2P_1\cdot l}{(P_1-x_2P_2+l)^2-m_b^2}
\frac{n_{+\beta}}{n_+\cdot l}
\frac{\not\! P_1+m_b}{(P_1-x_2P_2)^2-m_b^2}\;.
\nonumber
\end{eqnarray}
The neglect of $\xi_2\not P_2$ and $x_2\not P_2$ in the first and second
propagators, respectively, is due to $\gamma_\beta=\gamma^+$ in the
collinear region. The integral $I^{(g)}$ reduces to
\begin{eqnarray}
I^{(g)}&\approx &\int\frac{d^4l}{(2\pi)^4}
\frac{2P_1\cdot l}{(P_1-x_2P_2+l)^2-m_b^2}
\left(\frac{-g^2C_F}{2}m_0\right)\frac{tr[\gamma_\mu
(\not\! P_1+m_b)\gamma^\nu b(P_1-k_1)
{\bar d}(k_1)\gamma_\nu\gamma^5]}
{[(P_1-x_2P_2)^2-m_b^2](k_1-x_2P_2)^2}
\nonumber\\
& &\times \frac{-ig^2C_F}{4m_0}
u({\bar x}_2P_2)\gamma^\beta
\frac{{\bar x}_2\not\! P_2 +\not l}{({\bar x}_2P_2 +l)^2}\gamma_5
{\bar d}(x_2P_2) \frac{1}{l^2}
\frac{n_{+\beta}}{n_+\cdot l}\;.
\label{2bg}
\end{eqnarray}
The delicate combination of the first term in Eq.~(\ref{2bdf}) and
Eq.~(\ref{2bg}) leads to the correct factorization form with the
hard amplitude $H_{S}^{(1)}(x_1,\xi_2)$ and the collinear piece
proportional to $\delta (\xi_2-x_2)$:
\begin{eqnarray}
& &\int d\xi_2 H^{(0)}_{S}(x_1,\xi_2) \left[\frac{-ig^2C_F}{4m_0}
\int\frac{d^4l}{(2\pi)^4} u({\bar x}_2P_2)\gamma^\beta \frac{{\bar
x}_2\not\! P_2 +\not l}{({\bar x}_2P_2 +l)^2}\right.
\nonumber\\
& & \left.\;\;\times \gamma_5{\bar d}(x_2P_2) \frac{1}{l^2}
\frac{n_{+\beta}}{n_+\cdot
l}\delta\left(\xi_2-x_2\right)\right]\;.
\end{eqnarray}
The sum of Eqs.~(\ref{2bdf}) and (\ref{2bg}) then gives
Eq.~(\ref{2bdg}).

The loop integral from Fig.~2(h) is simplified to
\begin{eqnarray}
I^{(h)}\!&\approx&\!
\int d\xi_2 H^{(0)}_{S}(x_1,\xi_2)
\Bigg\{\frac{ig^2}{2m_0C_F}\int\frac{d^4l}{(2\pi)^4}
u({\bar x}_2P_2)\gamma_5
\frac{x_2\not\! P_2 -\not l}{(x_2 P_2 -l)^2}
\gamma^\beta{\bar d}(x_2P_2)\frac{1}{l^2}
\frac{n_{+\beta}}{n_+\cdot l}\delta(\xi_2-x_2)\Biggr\}
\nonumber\\
& &-\int\frac{d^4l}{(2\pi)^4}\Biggl(\frac{-g^2}{2}C_Fm_0\Biggr)
\frac{tr[\gamma_\mu (\not\! P_1-x_2\not\! P_2+m_b)\gamma^\alpha
b(P_1-k_1) {\bar d}(k_1)\gamma_\alpha\gamma^5]}
{[(P_1-x_2P_2)^2-m_b^2](k_1-x_2P_2+l)^2}
\nonumber\\
& & \times \frac{ig^2}{2m_0C_F} u({\bar x}_2P_2)\gamma_5
\frac{x_2\not P_2 -\not l}{(x_2 P_2 -l)^2} \gamma^\beta{\bar
d}(x_2P_2)\frac{1}{l^2} \frac{n_{+\beta}}{n_+\cdot l}\;.
\label{2bh}
\end{eqnarray}

The loop integral associated with Fig.~2(i) reduces to
\begin{eqnarray}
I^{(i)}&\approx & \int d\xi_2 H_{S}^{(0)}(x_1,\xi_2)
\phi_{Si}^{(1)}(x_2,\xi_2)\;,
\label{2bi}
\end{eqnarray}
with the collinear divergent piece $\phi_{Si}^{(1)}(x_2,\xi_2)$ shown in
Eq.~(\ref{p4bi}). Figure 2(j) leads to
\begin{eqnarray}
I^{(j)}& \approx &
\int\frac{d^4l}{(2\pi)^4}
\Bigg(\frac{-g^2}{2}C_Fm_0\Biggr)\frac{tr[\gamma_\mu
(\not\! P_1-x_2\not\! P_2+m_b)\gamma^\alpha b(P_1-k_1)
{\bar d}(k_1)\gamma_\alpha\gamma^5]}
{[(P_1-x_2P_2)^2-m_b^2](k_1-x_2P_2+l)^2}
\nonumber\\
& &\times
\frac{ig^2}{8m_0N_c}
u({\bar x}_2P_2)\gamma_5
\frac{x_2\not\! P_2-\not l}{(x_2P_2-l)^2}\gamma^\nu
{\bar d}(x_2P_2)\frac{1}{l^2}
\frac{n_{+\nu}}{n_+\cdot l}\;.
\label{2bj}
\end{eqnarray}
Combining Eqs.~(\ref{2bh}), (\ref{2bi}) and (\ref{2bj}),
we arrive at
\begin{eqnarray}
I^{(h)}\!+\!I^{(i)}\!+\!I^{(j)}
\!&\approx& \!
\int d\xi_2 H^{(0)}_{S}(x_1,\xi_2)
\Bigg\{\frac{ig^2}{4m_0C_F}\int\frac{d^4l}{(2\pi)^4}
u({\bar x}_2P_2)\gamma_5
\frac{x_2\not\! P_2 -\not l}{(x_2 P_2 -l)^2}
\gamma^\beta{\bar d}(x_2P_2)\frac{1}{l^2}
\frac{n_{+\beta}}{n_+\cdot l}\delta(\xi_2-x_2)\Biggr\}
\nonumber\\
& & -\int\frac{d^4l}{(2\pi)^4}
\Biggl(\frac{-g^2}{2}C_Fm_0\Biggr)\frac{tr[\gamma_\mu
(\not\! P_1-x_2\!\not\! P_2+m_b)\gamma^\alpha b(P_1-k_1)
{\bar d}(k_1)\gamma_\alpha\gamma^5]}
{[(P_1-x_2P_2)^2-m_b^2](k_1-x_2P_2+l)^2}\Bigg\}
\nonumber\\
& &\times \frac{ig^2}{4m_0C_F} u({\bar x}_2P_2)\gamma_5
\frac{x_2\not\! P_2 -\not l}{(x_2 P_2 -l)^2} \gamma^\beta{\bar
d}(x_2P_2)\frac{1}{l^2} \frac{n_{+\beta}}{n_+\cdot l}\;.
\label{2bhj}
\end{eqnarray}

The integral associated with Fig.~2(k) becomes
\begin{eqnarray}
I^{(k)}&\approx& \int\frac{d^4l}{(2\pi)^4}
\frac{2P_1\cdot l}{(P_1-x_2P_2)^2-m_b^2}
\left(\frac{-g^2}{2}C_Fm_0\right)\frac{tr[\gamma_\mu
(\not\! P_1+m_b)\gamma^\alpha b(P_1-k_1)
{\bar d}(k_1)\gamma_\alpha\gamma^5]}
{[(P_1-x_2P_2+l)^2-m_b^2](k_1-x_2P_2+l)^2}
\nonumber\\
& &\times\frac{ig^2}{4m_0C_F} u({\bar x}_2P_2)\gamma_5
\frac{x_2\not\! P_2 -\not l}{(x_2 P_2 -l)^2} \gamma^\beta{\bar
d}(x_2P_2)\frac{1}{l^2} \frac{n_{+\beta}}{n_+\cdot l}\;.
\label{2bk}
\end{eqnarray}
Combining the second term in Eq.~(\ref{2bhj}) and Eq.~(\ref{2bk}),
we obtain the correct factorization form in terms of the hard
amplitude $H_{S}^{(0)}(x_1,\xi_2)$ and the collinear piece
proportional to $\delta(\xi_2-x_2+l^-/P_2^-)$:
\begin{eqnarray}
& -& \int d\xi_2 H^{(1)}_{S}(x_1,\xi_2)
\left[\frac{ig^2}{4m_0C_F}\int\frac{d^4l}{(2\pi)^4} u({\bar
x}_2P_2)\gamma_5 \frac{x_2\not\! P_2 -\not l}{(x_2 P_2 -l)^2}\right.
\nonumber\\
& &\left.\times \gamma^\beta{\bar d}(x_2P_2)\frac{1}{l^2}
\frac{n_{+\beta}}{n_+\cdot
l}\delta\Biggl(\xi_2-x_2+\frac{l^-}{P_2^-}\Biggr)\right]\;.
\end{eqnarray}
The sum of Eqs.~(\ref{2bhj}) and (\ref{2bk}) then gives
Eq.~(\ref{2bhk}).

\newpage

{\bf \Large Figure Captions}
\vspace{10mm}

\begin{enumerate}

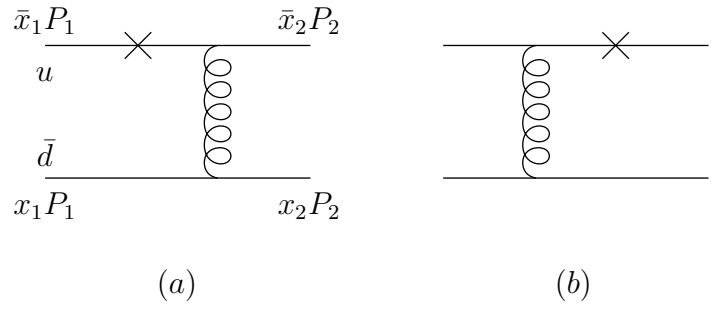
\item Fig. 1: Lowest-order diagrams for $\pi\gamma^*\to\pi$
($B\to\pi l\bar\nu$), where the symbol $\times$ represents the
virtual photon (weak decay) vertex.

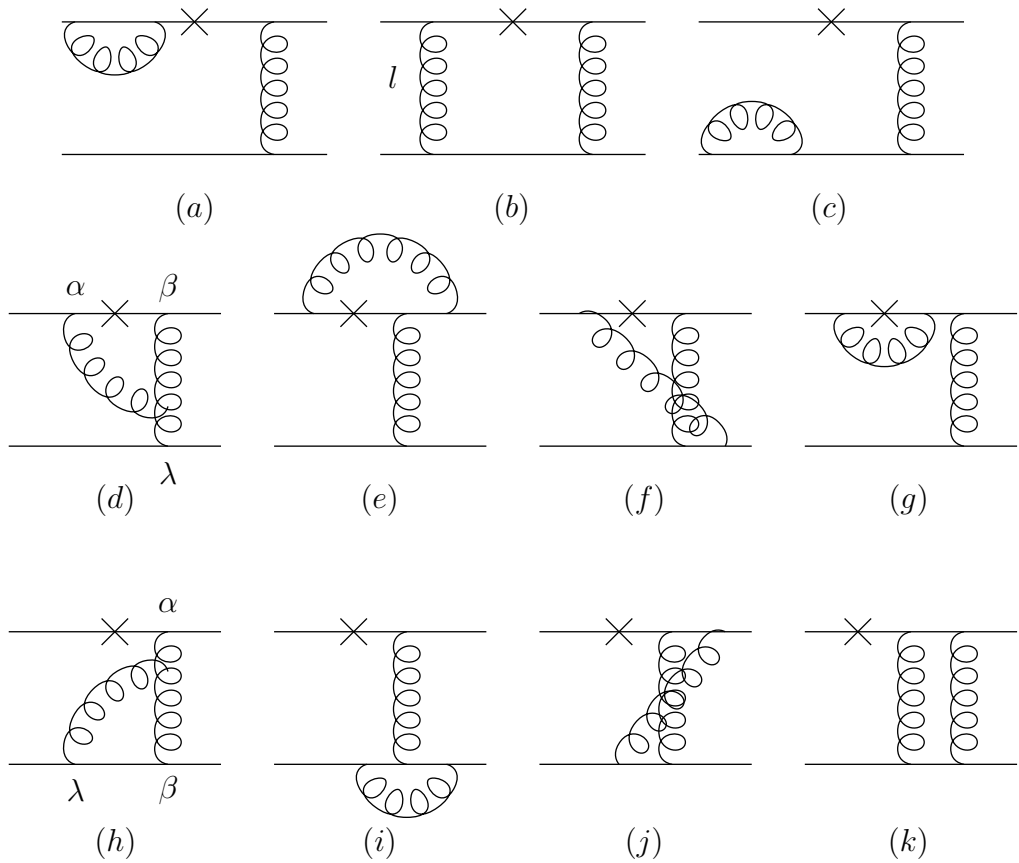
\item Fig. 2: $O(\alpha_s)$ radiative corrections to Fig.~1(a).

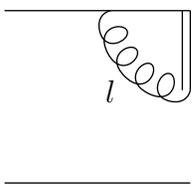
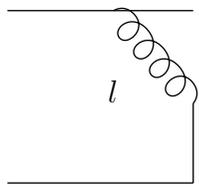
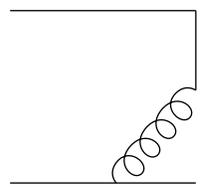
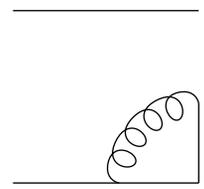
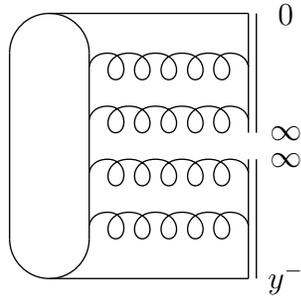
\item Fig. 3: (a)-(d) Infrared divergent diagrams factored out of
Fig.~2(d)-2(k). (e) The graphic definition of the two-parton twist-3 pion
distribution amplitudes.

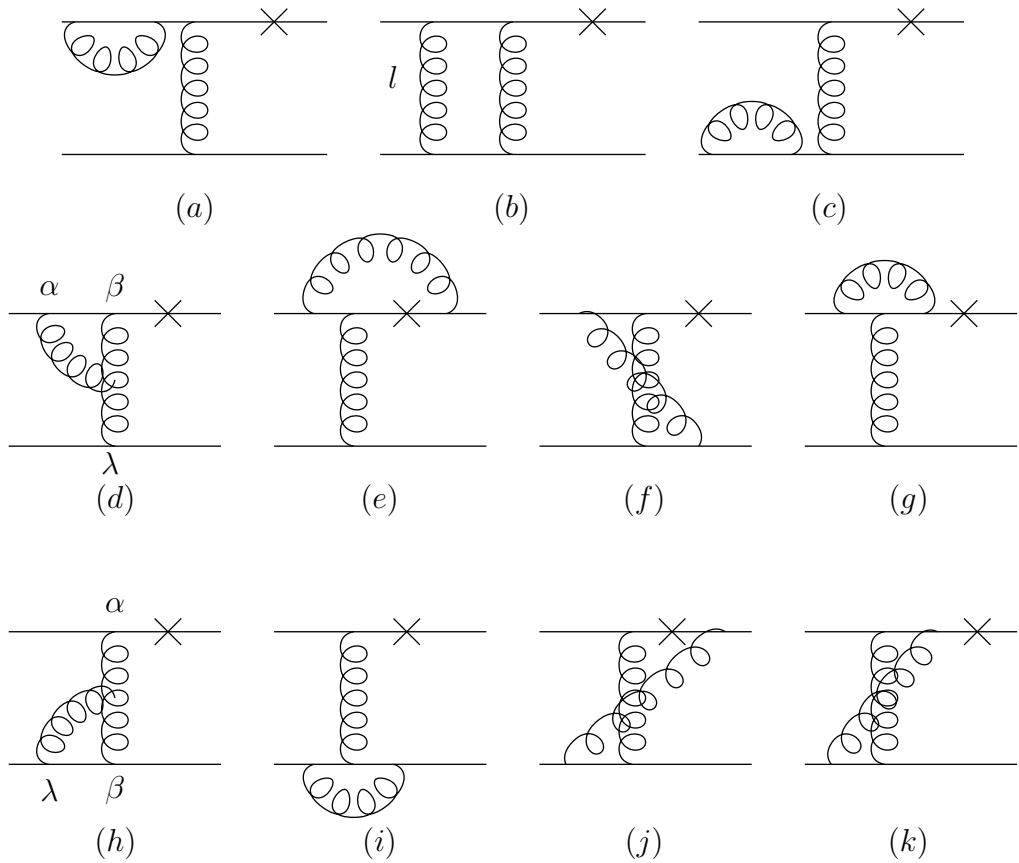
\item Fig. 4: $O(\alpha_s)$ radiative corrections to Fig.~1(b).

\item Fig. 5: (a) The Ward identity. (b) Factorization of
$O(\alpha_s^{N+1})$ diagrams as a result of (a).

\item Fig. 6: Factorization of $O(\alpha_s^{N+1})$ diagrams corresponding
to Eq.~(\ref{rep}).

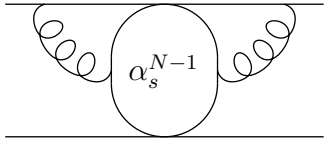
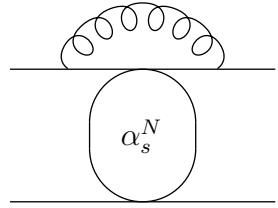
\item Fig. 7: (a) A typical diagram of
$G_{\parallel\;\parallel}^{(N+1)}$. (b) This diagram does not
belong to $G_{\parallel\;\parallel}^{(N+1)}$.

\end{enumerate}

\end{document}